\documentclass[11pt]{article}

\usepackage{amsmath}
\usepackage{amsthm}
\usepackage{amssymb}
\usepackage{amscd}
\usepackage{color}
\usepackage{soul}

\theoremstyle{definition}
\newtheorem{theorem}{Theorem}[section]
\newtheorem{prop}[theorem]{Proposition}
\newtheorem{lemma}[theorem]{Lemma}

\newtheorem{definition}[theorem]{Definition}
\newtheorem{example}[theorem]{Example}
\newtheorem{remark}[theorem]{Remark}

\numberwithin{equation}{section}
\newcommand\Nat{\mathbb{N}}
\newcommand\Int{\mathbb{Z}}

\newcommand\Comp{\mathbb{C}}
\newcommand\Real{\mathbb{R}}
\newcommand{\rmd}{\mathrm{d}}

\newcommand\bea{\begin{eqnarray}}
\newcommand\ena{\end{eqnarray}}
\newcommand\non{\nonumber}
\newcommand\tf{{\tilde f}}

\renewcommand\hat{\widehat}
\renewcommand\tilde{\widetilde}

\newcommand\UGM{\operatorname{UGM}}

\allowdisplaybreaks

\title
{
Sato Grassmannian and Degenerate Sigma Function  \\
}

\author{
Julia Bernatska\footnote{
National University of Kyiv-Mohyla Academy,
H. Skovorody St. 2, 04655 Kyiv, Ukraine,
{\tt jbernatska@gmail.com}
}
,\ 
Victor Enolski\footnote{
National University of Kyiv-Mohyla Academy,
H. Skovorody St. 2, 04655 Kyiv, Ukraine,
{\tt venolski@googlemail.com}
}
,\ 
and 
Atsushi Nakayashiki\footnote{
Department of Mathematics, Tsuda University,
Kodaira, Tokyo 187-8577, Japan,
{\tt atsushi@tsuda.ac.jp}
}
}

\date{
}

\begin{document}

\maketitle

\begin{flushright}
\centerline{\it Dedicated to Mikio Sato on his 90th birthday}
\end{flushright}

\begin{abstract}
The degeneration of  the hyperelliptic sigma function is studied.
We use the Sato Grassmannian for this purpose. A simple decomposition 
of a rational function gives a decomposition of Pl\"ucker coordinates of a frame  of the Sato Grassmannian.
It then gives a decomposition of the tau function corresponding to the degeneration 
of a hyperelliptic 
curve of genus $g$ in terms of the tau functions corresponding to a hyperelliptic curve of genus $g-1$.
Since the tau functions are described by sigma functions, we get the corresponding formula 
for the degenerate hyperelliptic sigma function.

\end{abstract}

\section{%
Introduction
}

We study an algebraic curve of genus $g>1$ and the associated 
   sigma-function of $g$ variables
 which represents a natural generalization 
of the Weierstrass elliptic sigma function to curves with higher genus \cite{BEL1,BEL2,KS,N2016}.
Although this multi-variate sigma function is defined as a certain modification of the Riemann 
theta function (see Definition \ref{def-sigma}), it has several nice properties which are absent in Riemann's theta 
function. The most important one is the modular invariance which means that it is invariant under the change 
of a canonical homology basis defining period matrices. 
For some class of algebraic curves such as $(n,s)$-curves a more strong property holds.
Namely the sigma function can be expanded to a power series whose coefficients are polynomials of the 
coefficients of the defining equation of the curve \cite{N2010-a,N2010,Ayano}. 
This property, in particular, implies that the sigma function has a well defined limit according as the limit of the 
defining equation of the curve.
Consequently the study of the limits of the sigma functions appears as an interesting and challenging 
problem. We remark that the period matrix defining the Riemann theta function diverges in general when 
a curve degenerates to a singular curve and the setting of the degeneration problem itself is non-trivial in this case \cite{BBEIM,Fay73}.

The degeneration problem  also attracts much attention in the study of integrable systems.
Solutions of KP type equations are built of
the sigma functions by known rules \cite{Kr1977,EEG,N2010}.
In particular they can be given 
as  logarithmic derivatives of the sigma-function.  Degenerate solutions 
appear when the curves have multiple points, in particular, double 
points.  A  degenerate  genus $g$ solution may be understood
as a {\em soliton on a finite-gap background} 
expressed in terms of the hyperbolic function  and the 
sigma function (see Chapter 4 of \cite{BBEIM} for the genus one background). 
Such solutions are in great demands 
of applied sciences and explicit formulae in closed form are requested (see \cite{KAA,GS,FLT}
and references therein).

In this paper we study the degeneration of the sigma function when a hyperelliptic curve of 
genus $g$ degenerates to a curve of genus $g-1$ with a double point.
To this end we study the degenerations of tau functions of the KP-hierarchy.
By using the Sato theory on the KP-hierarchy \cite{SS} the problem on tau functions reduces to 
a more easy problem on the frames of the Sato Grassmannian.
In this way we first derive the formula of the degenerate tau function.
Since the tau functions are expressed by the sigma functions,
the formula for the degenerate sigma function follows from that for the tau function.
This is the main idea of this paper.

Here we deal with the case of hyperelliptic curves.
But the method first manifested in \cite{N2018-1} on the genus zero degeneration 
of trigonal curves has a wider area of applicability. 
Thus  we consider the higher genus degeneration of hyperelliptic curve as a good laboratory for further generalizations.
We explain the main features of the method in more detail in the following.

First we show that the Sato theory, commonly 
considered as a deep theoretical description of integrability, can be implemented 
for the derivation of explicit formulae in a very concrete problem --- the construction of 
degenerate solutions in an explicit and closed form. In the case considered 
the hyperellptic curve with a double point leads to the decomposition of the 
Grassmannian frame and the decomposition  of  the tau function of genus $g$ 
to the tau functions of genus $g-1$ (Theorem \ref{tau-degeneration-rel}).

Next, we incorporate the viewpoint of \cite{BEL1} and further developments, in particular,
\cite{O2005,BL05,EEMOP07, N2010-a,NY2012,N2016,BEL2,MP,AN, BEL2018}, that the sigma function 
provides an algebraic counterpart of the Riemann theta function and 
an effective language to deal with the algebro-geometric integrability of soliton equations. 
After \cite{N2010,EH10,EEG}  the tau function is expressed  in terms of the sigma function  
and the interrelation of sigma and tau is a part of the construction. 
To derive the sigma function formula of the tau function  corresponding precisely
to a point of the Sato Grassmannian some detailed consideration is necessary.
For some tau function which appears in  the relation in Theorem \ref{tau-degeneration-rel} 
the formula is already known \cite{N2010} (see Theorem \ref{tau-sigma-1}). 
We derive the sigma function formula for the remaining tau functions in the relation
 (see Theorem  \ref{tau-Y}).

Also our derivation is based on the notion of Schottky-Klein bidifferential
whose fundamental role in the construction of Abelian function associated with
Riemann surfaces was elucidated in \cite{Fay73}.
Later, in \cite{SW} a
remarkable formula for the tau function corresponding to points of a Grassmannian
associated with algebro-geometric data was discovered. It is expressed by the Riemann
theta function and the normalized bidifferential.

In case a point of the Grassmannian is given algebraically in terms of the defining equation
of an algebraic curve, it is more natural to realize the tau function in algebraic form using
the sigma function. In such expression the normalized bidifferential is replaced by a non-normalized
algebraic bidifferential which is expressed in terms of curve coordinates \cite{N2010,EEG}.
For hyperelliptic curves
such representation of the bidifferential is known from H.Baker's
monograph \cite{Bak98} and nowadays it is obtained for wider
classes of curves \cite{BEL00,N2010-a,Suz17,Eyn18}.
As a consequence of this rewriting of the tau function, the sigma function
is conversely expressed using the tau function and algebraic quantities.
We use it to resolve the posted degeneration problem.

The derivation of the formula for the degenerate sigma function finally reduces
to calculating the difference of bidifferentials corresponding to the
degenerate limit of a genus $g$ curve and the
 non-singular  curve of genus $g-1$. We demonstrate that, with the help of
the explicit algebraic expression of the bidifferential,
 it can neatly be  expressed  in terms of the coefficients of the
defining equation of the curve. In this way we get the formula which
expresses the degenerate sigma function of genus $g$
 in terms of the sigma functions of genus $g-1$ (Theorem \ref{main}).
The case of genus two was previously
 studied in \cite{BL}, where the system of linear differential equations
satisfied by the sigma function \cite{BL05} was used to derive the formula.
Our result generalizes it to the case of genus $g$ by another method
(see section 6 for details).


Here it should be emphasized that the method proposed here permits 
to avoid  calculations  of both
limits of the Riemann matrix and limits of second kind transcendents 
entering to the definition of sigma. Then the whole procedure reduces to 
routine expansions of algebraic quantities such as the  non-normalized
bidifferential represented in algebraic form.
Thus we can expect that it is effective for the more complicated curves 
than hyperelliptic curves.

Recently degenerations of curves to genus $g=0$ and associated trigonometric solutions of the KP 
hierarchies were studied in \cite{Kod17,AG18} (see also references therein). It is shown that 
the Sato Grassmannian approach is effective in this case \cite{N2018-1,N2018-2}.
 We believe that our consideration will permit us to extend 
  interesting results and observation found in these publications to the case of  genus $g>1$ degeneration of   a curve.

The organization of the paper is as follows.
In section 2 we recall the definition  of the Sato Grassmannian, which we denote by UGM, 
 and the correspondence between solutions (tau functions) of the KP-hierarchy and points 
of UGM.
We define points of UGM from some spaces of  functions on  hyperelliptic curves and consider the 
corresponding tau functions in section 3. We introduce frames $\tilde{X}$, $X$, $Y^\pm$.
The frame $\tilde{X}$ is associated with a non-singular curve of genus $g$ and $X$ is 
its degenerate limit. The frames $Y^\pm$ correspond to the curve of genus $g-1$.
By examining the description of frames we derive the expression of the tau function of $X$ 
as a linear combination of the tau functions of $Y^\pm$.
In section 4 we first recall the definition of the fundamental sigma function, which is simply called the 
sigma function in the other part of the paper, and the sigma function expression of the tau function 
of $\tilde{X}$ given in \cite{N2010}. Then we derive the formulas of the tau functions for $Y^\pm$ 
in terms of the sigma function. For the derivation we use several properties of the sigma 
function \cite{O2005,NY2012,N2016}.
By substituting the sigma function formulas of tau functions to the relation derived in 
section 3, the formula for the degenerate sigma function of genus $g$ in terms of 
the sigma function of genus $g-1$ is derived in section 5. 
The variables of the tau function have a universal character which means that 
they do not depend on the genus of curves. However the variables of the sigma function 
depend on the genus of the curve. To connect the variables of the sigma functions of different genus 
and to express the quantities in the formula for the degenerate sigma functions we examine 
the expansions of the holomorphic differentials and the meromorphic bidifferentials.
In section 6,  we show that the formula of \cite{BL} for the degenerate
sigma function of genus two can be derived by the method proposed in this paper.

\section{%
Sato Grassmannian and tau function
}
In this section we recall the definition of the Sato Grassmannian, which we denote by UGM 
(universal Grassmann manifold) after M.Sato, and the correspondence between solutions (tau functions) 
of the KP-hierarchy and points of UGM \cite{SS}.

In the following we denote by $\Int$ the set of integers and by $\Nat$ the set of positive integers.

Let 
\bea
&&
V={\mathbb C}((z))=\bigg\{\sum_{-\infty \ll i<+\infty} a_i z^i\, |\, a_i\in {\mathbb C}\bigg\}
\non
\ena
be the vector space of Laurent series in $z$ and 
\bea
&&
V_\emptyset={\mathbb C}[z^{-1}],
\quad
V_0=z{\mathbb C}[[z]].
\non
\ena
Then 
$$
V = V_\emptyset \oplus V_0.
$$
Define the projection map from $V$ to $V_\emptyset$ by $\pi$:
\bea
&&
\pi: V \rightarrow V/V_0\simeq V_\emptyset.
\non
\ena

The Sato Grassmannian UGM is defined as the set of subspaces $U$ of $V$ such that 
${\rm Ker}(\pi|_U)$ and ${\rm Coker} (\pi|_U)$ are finite dimensional and their dimensions are 
equal.

\begin{example}
$U=V_\emptyset\in \UGM$. In this case $\dim {\rm Ker} (\pi|U)=\dim {\rm Coker} (\pi|U)=0$.
\end{example}

\begin{example} 
$\displaystyle{U=\Comp z+\sum_{i=0}^\infty \Comp z\wp^{(i)}(z) \in \UGM}$, where $\wp(z)$ is the Weierstrass elliptic function and 
$\wp^{(i)}(z)= \rmd^i \wp(z)/\rmd z^i$.  In this case $\dim {\rm Ker} (\pi|U)=\dim {\rm Coker} (\pi|U)=1$.
\end{example}

We set 
\bea
&&
e_i=z^{i+1}.
\non
\ena
A basis $\xi=(\ldots,\upsilon_3,\upsilon_2,\upsilon_1)$ of $U\in \UGM$ is called a frame of $U$.
Write 
\bea
\upsilon_j=\sum_{-\infty\ll i<+\infty}\xi_{i,j} e_i
\non
\ena
 and identify $\upsilon_j$ with the column vector $(\xi_{i,j})_{i\in \Int}$,
 \bea
 &&
 \upsilon_j=\left[
 \begin{array}{c}
 \vdots\\
 \xi_{-1,j}\\
 \xi_{0,j}\\
 \xi_{1,j}\\
 \vdots\\
 \end{array}
 \right].
 \non
 \ena
 Then $\xi$ can be expressed as a $\Int\times \Nat$ matrix $(\xi_{i,j})_{i\in \Int,j\in \Nat}$.
 
For any point $U$ of UGM it is always possible to take a frame in the following form:
there exists $l\geq 1$ such that, for $j\geq l$,
\bea
&&
\upsilon_j=e_{-j}+\sum_{i>-j}\xi_{i,j}e_i.
\label{frame-normal}
\ena In the matrix form it means that the frame has the following form
 
 \bea
&&
\xi=\left[\begin{array}{cccccc}
\ddots&O&\vdots&\vdots&\quad&\vdots\\
\cdots&1&0&\ast&\cdots&\ast\\
\cdots&\ast&1&\ast&\cdots&\ast\\
\vdots&\vdots&\vdots&\vdots&\quad&\vdots\\
\end{array}
\right],
\label{frame}
\ena
where the right most $1$ in (\ref{frame}) is the $(-l,l)$ component.
In the following we assume that a frame is always of this form.
 
 A sequence of integers $M=(m_1,m_2,\ldots)$ such that 
 \bea
 &&
 m_1>m_2>\cdots,
 \non
 \\
 &&
 m_j=-j \quad \forall j>>0,
 \non
 \ena
 is called a Maya diagram. 
The set of Maya diagrams and that of partitions bijectively correspond to each other 
by
\bea
M=(m_1,m_2,\ldots) &\longrightarrow& \lambda(M)=(m_1+1,m_2+2,\ldots),
\non
\\
\lambda=(\lambda_1,\lambda_2,\ldots) &\longrightarrow& M(\lambda)=(\lambda_1-1,\lambda_2-2,\ldots).
\non
\ena
For a partition $\lambda$ the Pl\"ucker coordinate $\xi_\lambda$ of a frame $\xi$ is defined 
by
\bea
&&
\xi_\lambda=\det(\xi_{m_i,j})_{i,j\in \Nat},
\non
\\
&&
M(\lambda)=(m_1,m_2,\ldots).
\non
\ena
The infinite determinant is well defined for a frame of the form (\ref{frame}).

The Schur function $s_{(i)}(t)$, $t=(t_1,t_2,\ldots)^t$, 
corresponding to the partition $(i)$,  $i\geq 0$, is defined by 
\bea
&&
\exp \bigg(\sum_{i=1}^\infty t_i k^i\bigg)=\sum_{i=0}^\infty s_{(i)}(t) k^i.
\non
\ena
For an arbitrary partition $\lambda=(\lambda_1,\lambda_2,\ldots, \lambda_l)$ 
the Schur function $s_{\lambda}(t)$ is defined by
\bea
&&
s_\lambda(t)=\det\left(s_{(\lambda_i-i+j)}(t)\right)_{1\leq i,j\leq l}.
\non
\ena
Assign the weight $i$ for $t_i$. Then $s_\lambda(t)$ is homogeneous of the weight $|\lambda|=\sum \lambda_i$.

The tau function of a frame $\xi$ of a point $U$ of $\UGM$ is defined by
\bea
&&
\tau(t;\xi)=\sum_\lambda \xi_\lambda s_\lambda(t),
\label{tau-series}
\ena
where the sum is over all partitions.
If $\xi'$ is another frame of $U$ then $\tau(t;\xi')$ is a constant multiple of $\tau(t;\xi)$.

The tau function gives a solution of the KP-hierarchy.
The KP-hierarchy in the bilinear form is the equation for $\tau(t)$, $t=(t_1,t_2,\dots)^t$ given by
\bea
&&
\oint {\rm e}^{-2\sum_{j=1}^\infty s_j z^j}\tau\big(t-s-[z^{-1}]\big)
\tau\big(t+s+[z^{-1}]\big)\frac{\rmd z}{2\pi i}=0,
\label{KP-hierarchy}
\ena
where $s=(s_1,s_2,s_3, \dots)^t$, $[z]=(z,z^2/2,z^3/3,\dots)$ 
and the integral signifies to take the coefficient of $z^{-1}$ in the formal series expansion of the integrand in $z$.

For a solution $\tau(t)$ of the KP-hierarchy  $u(t)=2\partial_{t_1}^2\log \tau(t)$
 satisfies the KP equation 
\bea
&&
3u_{t_2t_2}+(-4u_{t_3}+6uu_{t_1}+u_{t_1t_1t_1})_{t_1}=0.
\label{KP-equation}
\ena

Sato's fundamental theorem tells that any formal power 
series solutions of the KP-hierarchy can be constructed from UGM.

\begin{theorem}\label{Sato1}{\rm \cite{SS}} For any $U\in \UGM$ and a frame $\xi$ of $U$ $\tau(t;\xi)$ 
is a solution of the KP-hierarchy. Conversely for any formal 
power series solution $\tau(t)$ of the
KP-hierarchy there exists a unique point $U$ of ${\rm UGM}$ and a frame $\xi$ of $U$ such that
$\tau(t)=\tau(t;\xi)$.
\end{theorem}

The point $U$ of UGM corresponding to a solution $\tau(t)$ in Theorem \ref{Sato1} is 
constructed in the following manner \cite{SS,KNTY,N2010}.

Let $\Psi^\ast(t;z)$ be the adjoint wave function \cite{DJKM} corresponding to $\tau(t)$ 
which is defined by
\bea
&&
\Psi^\ast(t;z)=\frac{\tau(t+[z])}{\tau(t)}\exp(-\sum_{i=1}^\infty t_iz^{-i}).
\label{wave}
\ena
Define $\Psi_i^\ast(z)$ by the following expansion
\bea
\left(\tau(t)\Psi^\ast(t;z)\right)\vert_{t=(x,0,0,0,...)}
&=&
\tau((x,0,0,0,...)+[z])\exp(-x z^{-1})
\non
\\
&=&
\sum_{i=0}^\infty \Psi_i^\ast(z) x^i.
\label{wave-expand}
\ena
Then
\bea
&&
U=\sum_{i=0}^\infty {\mathbb C}\Psi_i^\ast(z).
\label{U-tau}
\ena

\section{Tau functions of hyperelliptic curves}
In this section we describe frames of UGM corresponding to some spaces of 
 functions on hyperelliptic curves.  Using them we derive the expression of the degenerate limit of a 
 tau function of a hyperelliptic curve of genus $g$  in terms of tau functions of a hyperelliptic curve of 
 genus $g-1$.
 
\subsection{The case of zero point}
Here we construct a frame of UGM from a hyperelliptic curve of genus $g$.

Consider the hyperelliptic curve $C$ of genus $g$ given by
\bea
&& y^2=\prod_{j=1}^{2g+1}(x-\alpha_j),
\label{curve-1}
\ena
where  $\{\alpha_i\}$ are mutually distinct complex numbers.
The space of meromorphic functions on $C$ which are regular on $C\backslash\{\infty\}$ is denoted by $H^0(C,{\cal O}(\ast \infty))$. We set
\bea
&&
\tilde{W}=H^0(C,{\cal O}(\ast \infty)).
\label{def-W}
\ena

A basis of $\tilde{W}$ is given by
\bea
&&
x^i , \quad i\geq 0,
\hskip5mm
x^iy, \quad i\geq 0.
\label{basis-1}
\ena
We take the local coordinate $z$ around $\infty$ such that 
\bea
&&
x=z^{-2},
\hskip5mm
y=z^{-(2g+1)}F(z), 
\hskip5mm
F(z)=\left(\prod_{j=1}^{2g+1}(1-\alpha_j z^2)\right)^{1/2},
\label{local-coord}
\ena
where $F(z)$ is considerer as a power series in $z$ by the Taylor expansion at $z=0$.
By expanding functions in $z$ we consider $\tilde{W}$ as a subspace of $V={\mathbb C}((z))$. 
Then (see Corollary 1 of \cite{N2018-2})
\bea
&&
z^g\tilde{W} \in {\rm UGM}.
\label{zg-tildeW}
\ena
Writing the basis (\ref{basis-1}) in terms of $z$ and multiplying them by $z^g$ we get a basis of 
$z^g\tilde{W}$,
\bea
&&
z^{g-2i}, \quad i\geq 0, 
\hskip5mm
z^{-g-1-2i}F(z), \quad i\geq 0.
\label{basis-3}
\ena
If a Laurent series $v(z)$ is of the form
\bea
&&
v(z)=\sum_{n\geq N} a_n z^n,
\hskip5mm
a_{N}\neq 0,
\non
\ena
we define the order of $v(z)$ to be $-N$. 
Arrange elements of (\ref{basis-3}) by their orders and denote them by $\tf_1,\tf_2,...$.
Explicitly
\bea
(\tf_1,...,\tf_g)&=&(z^g,z^{g-2},z^{g-4},...,z^{-g+2}),
\non
\\
\tf_{g+2i+1}&=&z^{-g-2i}, \quad i\geq 0
\non
\\
\tf_{g+2i+2}&=& z^{-g-1-2i}F(z), \quad i\geq 0.
\non
\ena
Define the frame $\tilde{X}$ of a point of UGM by
\bea
&&
\tilde{X}=[...,\tf_3,\tf_2,\tf_1].
\label{frame-tilde-X}
\ena
Notice that $\tilde{X}$ satisfies (\ref{frame-normal}).

\subsection{Degeneration}
We study the degenerations of the curve $C$ and the frame $\tilde{X}$.

Let $\alpha$ be a complex number which is different from $\alpha_j$, $1\leq j\leq 2g-1$.
We consider the limit of $C$ when $\alpha_{2g}$ and $\alpha_{2g+1}$ go to $\alpha$. 
The curve $C$ tends to
\bea
&&
y^2=(x-\alpha)^2\prod_{j=1}^{2g-1}(x-\alpha_j).
\label{curve-2}
\ena
The limit of the function $F(z)$ becomes
\bea
&&
F(z)=(1-\alpha z^2)F_0(z),
\hskip5mm
F_0(z)=\left(\prod_{j=1}^{2g-1}(1-\alpha_j z^2)\right)^{1/2},
\non
\ena
where $F_0(z)$ is again considered as a series in $z$.
Then the limit of  $\tf_j$ exists and it is obtained by replacing $F(z)$ by $(1-\alpha z^2)F_0(z)$ in  
$\tf_j$. Denote the limit of  $\tf_j$ by $\tf_j^0$.
Define   ${\tilde X}_0$ as the limit of the frame $\tilde{X}$,
\bea
&&
{\tilde X}_0=[...,\tf_3^0,\tf_2^0,\tf_1^0].
\non
\ena
It still is a frame of a point of a UGM, since the highest order terms of $f_j$ and $\tilde{f}_j^0$ are the same.
Let $\tilde{U}_0$ be the corresponding point of UGM.
 
We multiply ${\tilde X}_0$ by $(1-\alpha z^2)^{-1}$. Then the following set of functions
is a basis of $(1-\alpha z^2)^{-1}\tilde{U}_0$,
\bea
&&
\frac{z^{g-2i}}{1-\alpha z^2},\quad i\geq 0,
\hskip5mm
z^{-g-1-2i} F_0(z), \quad i\geq 0.
\label{dbasis-1}
\ena

\begin{lemma}\label{basis-change-f0}
We have the following equation
\bea
\left<\frac{z^{g-2i}}{1-\alpha z^2}, \quad i\geq 0\right>=
\left<\frac{z^{g}}{1-\alpha z^2}, z^{g-2-2i}, \quad i\geq 0 \right>,
\label{change-f0}
\ena
where, for a set of functions $S$, $<S>$ denotes the vector space generated by $S$.
\end{lemma}
\vskip2mm
\noindent
{\it Proof.} 
Notice that $z^g/(1-\alpha z^2)$ belongs to both the LHS and the RHS of the equation (\ref{change-f0}).
Then using the relation
\bea
&&
\frac{z^{g-2i}-\alpha z^{g-2(i-1)}}{1-\alpha z^2}=z^{g-2i}
\label{fundamental-rel}
\ena
it is easy to show that the RHS is included in the LHS by induction on $i$.
The converse inclusion relation can be proved by using 
\bea
&&
\frac{z^{g-2(i+1)}}{1-\alpha z^2}=z^{g-2(i+1)}+\frac{\alpha  z^{g-2i}}{1-\alpha z^2}.
\non
\ena
$\Box$

By Lemma \ref{basis-change-f0} and  (\ref{dbasis-1}) the set of functions
\bea
&&
\frac{z^{g}}{1-\alpha z^2},
\hskip5mm
 z^{g-2-2i}\,\, (i\geq 0),
\hskip5mm
z^{-g-1-2i}F_0(z) \,\, (i\geq 0).
\label{dbasis-2}
\ena
becomes a basis of $(1-\alpha z^2)^{-1}\tilde{U}_0$.
Arrange the functions in (\ref{dbasis-2}) by their orders and denote them by $f_1,f_2,...$,
\bea
(f_1,f_2,f_3,...,f_g)&=&
\left(\frac{z^{g}}{1-\alpha z^2}, z^{g-2},z^{g-4},...,z^{-g+2}\right),
\nonumber
\\
f_{g+2i+1}&=&z^{-g-2i}, \quad i\geq 0,
\nonumber
\\
f_{g+2i+2}&=&z^{-g-2i-1}F_0(z), \quad i\geq 0.
\non
\ena
Define the frame $X$ of UGM by
\bea
&&
X=(X_{i,j})_{i\in {\mathbb Z}, j\in \Nat}=[...,f_3,f_2,f_1].
\label{gauge-trf}
\ena
In the change from (\ref{dbasis-1}) to (\ref{dbasis-2}) only the relation (\ref{fundamental-rel}) 
is used. It does not change the Pl\"ucker coordinates. Therefore the tau functions of $X$ and 
$(1-\alpha z^2)^{-1}{\tilde X}_0$ are the same,
\bea
&&
\tau(t;X)=\tau\big(t;(1-\alpha z^2)^{-1}\tilde{X}_0\big).
\label{tau-X}
\ena

\subsection{The case of one point}
In this section we study the frames of UGM corresponding to the space of functions 
on a hyperelliptic curve of genus $g-1$ which have at most a simple pole at some fixed point  and 
a pole at $\infty$ of arbitrary order.
We use them to describe the frame $X$.

Consider the hyperelliptic curve $C'$ of genus $g-1$ given by 
\bea
&&
y^2=\prod_{j=1}^{2g-1}(x-\alpha_j),
\label{curve-C-dash}
\ena
which is extracted from the singular curve (\ref{curve-2}).

Let $z$ be the local coordinate of $C'$ around $\infty$  such that
\bea
&&
x=z^{-2},
\hskip5mm
y=z^{-(2g+1)}F_0(z)
\non
\ena
and $\alpha$ be as in the previous section. 

Fix a point $(\alpha,y_0)$ of $C'$ and set 
\bea
&&
p_\pm=(\alpha,\pm y_0).
\label{p-pm}
\ena
Since $\alpha$ is different from $\alpha_j$, $1\leq j\leq 2g-1$, $y_0\neq 0$ and $p_+\neq p_-$.

For a point $P$ of $C'$ denote by $H^0(C',{\cal O}(P+\ast \infty))$ the space of meromorphic 
functions on $C'$ which are regular on $C'\backslash\{P,\infty\}$ and have a pole at $P$ of order at most one.
Define the vector spaces $W_\pm$ by
\bea
&&
W_\pm=H^0(C',{\cal O}(p_\pm+\ast \infty)).
\label{W-pm}
\ena
We consider $W_\pm$ as a subspace of $V={\mathbb C}((z))$ by expanding elements in $W\pm$ in $z$. 
Then (see Corollary 1 of  \cite{N2018-2})
\bea
&&
z^{g-2}W_\pm\in {\rm UGM}.
\label{UGM-pmY}
\ena

The following elements give a basis $z^{g-2}W_\pm$,
\bea
&&
z^{g-2-2i}, \quad i\geq 0,
\hskip5mm
z^{-g-1-2i} F_0(z), \quad i\geq 0, 
\hskip5mm
z^{-g+1} \frac{F_0(z)\pm y_0 z^{2g-1}}{1-\alpha z^2}.
\label{g-1basis-2}
\ena
Arrange the functions in (\ref{g-1basis-2}) by their orders and name them as $h^\pm_1,h^\pm_2,...$,
\bea
(h^\pm_1,h^\pm_2,...,h^\pm_g)&=&
\left(z^{g-2},z^{g-4},...,z^{-g+2}, z^{-g+1} \frac{F_0(z)\pm y_0 z^{2g-1}}{1-\alpha z^2}\right),
\nonumber
\\
h^\pm_{g+2i+1}&=&z^{-g-2i}, \quad i\geq 0,
\nonumber
\\
h^\pm_{g+2i+2}&=&z^{-g-1-2i} F_0(z), \quad i\geq 0.
\nonumber
\ena
We define the frames $Y^\pm$ of  points of UGM by
\bea
&&
Y^\pm=[...,h^\pm_3,h^\pm_2,h^\pm_1].
\label{frame-Y}
\ena

\subsection{Relation of tau functions}
We express the tau function corresponding to the frame $X$ in terms of the tau functions corresponding to 
the frames $Y^\pm$.

Since $h^+_i$ and $h^-_i$ coincide except $i=g$, we set, for $i\neq g$,
$$
h_i=h_i^+=h_i^- .
$$
Then $h_i$ is related to $f_i$ by
\bea
&&
h_i=\left\{\begin{array}{ll}
f_{i+1} & 1\leq i\leq g-1\\
f_i &i\geq g+1.
\end{array}
\right.
\ena
The following important decomposition holds,
\bea
&&
f_1=(2y_0)^{-1}(h^+_g-h^-_g).
\label{decomposition-f1}
\ena

Substituting  (\ref{decomposition-f1}) into the frame $X$, for any partition $\lambda$, the Pl\"ucker coordinate $X_\lambda$ 
can be written in terms of the Pl\"ucker coordinates $Y^\pm_\lambda$ as
\bea
&&
X_\lambda=(-1)^{g-1}(2y_0)^{-1}\left(Y^+_\lambda-Y^-_\lambda\right).
\label{decomposition-Plucker}
\ena
Thus we have 

\begin{theorem}\label{tau-degeneration-rel}
Let $C$ be the hyperelliptic curve of genus $g$ defined by (\ref{curve-1}) and $\tilde{X}$ 
the frame defined by (\ref{frame-tilde-X}) corresponding to  $C$.  Consider the degeneration 
of $C$ to the curve defined by (\ref{curve-2}) with one double point. Denote by $X$ the frame defined by 
 (\ref{gauge-trf}) which is obtained from the limit of $\tilde{X}$ and by $Y^\pm$ 
 the frames defined by (\ref{frame-Y}) corresponding to the non-singular curve of genus $g-1$ defined by (\ref{curve-C-dash}).

Then
\bea
&&
\tau(t;X)=(-1)^{g-1}(2y_0)^{-1}\left(\tau(t;Y^+)-\tau(t;Y^-)\right).
\label{decomposition-tau}
\ena
\end{theorem}

\section{Tau function in terms of sigma function}
In the previous section frames $\tilde{X}$, $\tilde{X}_0$, $Y^\pm$ of points of UGM are introduced.
The tau functions corresponding to those frames are defined as series as in (\ref{tau-series}).
There are analytic expressions of these tau functions in terms of multi-variable sigma functions.
In this section we give them.

\subsection{Hyperelliptic sigma function}
Sigma functions are defined for an arbitrary Riemann surface \cite{KS,N2016}. To recall the general case is too much for our purpose. On the other hand, in the case of hyperelliptic curves,  everything is explicit and is easy to understand. So we restrict ourselves to the hyperelliptic curve $C$ and recall the definition and properties 
of the sigma function of $C$.
To this end we need period matrices of certain first and second kind differentials \cite{BEL1, N2010-a}.

Write the right hand side of (\ref{curve-1}) as 
\bea
&&
\prod_{j=1}^{2g+1}(x-\alpha_j)=\sum_{i=0}^{2g+1} \mu_{4g+2-2i} x^i,
\hskip5mm
\mu_{0}=1.
\label{curve-RHS}
\ena
We assign the weight $k$ to $\mu_k$, and weight $2$ to $x$.
Then this polynomial is homogeneous of the weight $4g+2$.

Let $\{\rmd u_{2i-1}^{(g)}\,|\, 1\leq i\leq g\,\}$ be a basis of holomorphic one forms given by
\bea
&&
\rmd u_{2i-1}^{(g)}=\frac{x^{g-i} \rmd x}{-2y},
\label{abel1}
\ena
and $\widehat{\Omega}^{(g)}(p_1,p_2)$ the bidifferential on $C\times C$ defined by
\bea
\widehat{\Omega}^{(g)}(p_1,p_2) &=&
\frac{2y_1y_2+\sum_{i=0}^g x_1^i x_2^i (2\mu_{4g+2-4i}+\mu_{4g-4i}(x_1+x_2)) }
{4(x_1-x_2)^2 y_1y_2}\, \rmd x_1 \rmd x_2
\non
\\
&=& \rmd_{p_2} 
\left(\frac{y_1+y_2}{2y_1(x_1-x_2)}\right)+\sum_{i=1}^g \rmd u_{2i-1}^{(g)}(p_1) \rmd r_{2i-1}^{(g)}(p_2),
\label{bidiff1}
\ena
where $p_i=(x_i,y_i)\in C$ and 
\bea
&&
\rmd r_{2i-1}^{(g)}=\frac{\rmd x}{-2y}\sum_{k=1}^{2i-1}k\mu_{4i-2k-2} x^{k+g-i}.
\non
\ena
The differential $\rmd r_{2i-1}^{(g)}$ is a second kind differential with a pole only at $\infty$ and 
$\{\rmd u_{2i-1}^{(g)}, \rmd r_{2j-1}^{(g)}\}$ forms a symplectic basis of $H^1(C,{\mathbb C})$
with respect to the intersection form \cite{N2010}.

Take a canonical homology basis $\{\mathfrak{a}_i,\mathfrak{b}_i|1\leq i\leq g\}$ of the first 
homology group of $C$ and 
define the $g\times g$ period matrices by
\bea
&&
2\omega^{(g)}=\left(\int_{\mathfrak{a}_j} \rmd u_{2i-1}^{(g)}\right),
\hskip10mm
2\omega'{}^{(g)}=\left(\int_{\mathfrak{b}_j} \rmd u_{2i-1}^{(g)}\right),
\non
\\
&&
-2\eta^{(g)}=\left(\int_{\mathfrak{a}_j} \rmd r_{2i-1}^{(g)}\right),
\hskip10mm
-2\eta'{}^{(g)}=\left(\int_{\mathfrak{b}_j} \rmd r_{2i-1}^{(g)}\right),
\non
\\
&&
\quad\,\,
T^{(g)}= \big(\omega^{(g)}\big)^{-1}\omega'{}^{(g)}.
\ena

With the normalized period matrix $T^{(g)}$ the Riemann theta function with characteristics is 
defined by 
\bea
&&
\theta\begin{bmatrix} \varepsilon \\ \varepsilon' \end{bmatrix} (v|T^{(g)})
=\sum_{n\in\Int^g}\exp\left(\pi i (n+\varepsilon')^t T^{(g)}(n+\varepsilon')
+ 2\pi i (n+\varepsilon')^t (v + \varepsilon)\right),
\non \ena
where $\varepsilon,\varepsilon'\in \Real^g$.
To characteristics $\varepsilon,\varepsilon'\in \Real^g$ a vector $\vec{v}\in {\mathbb C}^g$ is 
associated by
\bea
&&
v=I^{(g)}\varepsilon +T^{(g)}\varepsilon',
\non
\ena
where $I^{(g)}$ is the unit matrix of degree $g$. This gives a bijection  
between the set of characteristics and the vectors in ${\mathbb C}^g$.
For a vector $v\in {\mathbb C}^g$ we denote $[v]$ the characteristics 
$\displaystyle \begin{bmatrix} \varepsilon \\ \varepsilon' \end{bmatrix}$.

Choose $\infty$ as a base point and denote by $K$ the Riemann's constant for this choice.  
Let $\displaystyle [-K]=\begin{bmatrix} \kappa \\ \kappa' \end{bmatrix}$.
It is known that $[-K]$ is a half period, that is, all components of $\kappa,\kappa'$ are half integers (see 
\cite{BEL1,Mum} for example).

In the definition of the sigma function we need some normalization constant which is described from
a certain derivative of  $\theta[-K]((2\omega^{(g)})^{-1} u | T^{(g)})$, 
$u=(u_1$, $u_3$, $\dots$, $u_{2g-1})^t$.
We recall necessary results concerning to it.

Set 
\bea
&&
m_k^{(g)}=\left[\frac{g+1-k}{2}\right], 
\quad
0\leq k\leq g,
\non
\\
&&
(w_1,...,w_g)=(1,3,5,...,2g-1),
\hskip5mm
(w_1^\ast,...,w_g^\ast)=(0,2,4,...,2g-2).
\non
\ena
For  $0\leq k\leq g-1$ define 
\bea
&&
A_k^{(g)}=(a^{(g)}_{k,1},...,a^{(g)}_{k,m_k^{(g)}})=(2g-2k-1,2g-2k-5,2g-2k-9,...),
\label{akg}
\\
&&
\mathrm{s}_k^{(g)}={\rm sgn}\left(
\begin{array}{cccccc}
w_1^\ast&...&w_{m_k^{(g)}}^\ast&w_{g-k-m_k^{(g)}}&...&w_1\\
g-k-1&\quad&...&...&1&0\\
\end{array}
\right).
\label{signature}
\ena
We set $A_g^{(g)}=\emptyset$, ${s}_g^{(g)}=1$. The following relation is valid,
\bea
&&
A^{(g)}_k=A^{(g-k)}_0.
\non
\ena

\begin{example} 
 $A^{(3)}_2=A^{(2)}_1=A_0^{(1)}=(1)$,
\par
\noindent
 $A^{(4)}_2=A^{(3)}_1=A_0^{(2)}=(3)$,
\par
\noindent
$A^{(5)}_2=A^{(4)}_1=A_0^{(3)}=(5,1)$,
\par
\noindent
$A^{(6)}_2=A^{(5)}_1=A_0^{(4)}=(7,3)$.
\end{example}
\vskip5mm

\begin{remark} The number $A^{(g)}_k$ was introduced in \cite{O2005} by different notation 
for hyperelliptic curves. Its generalizations are given in \cite{NY2012}  for $(n,s)$ curves
and in \cite{N2016} for arbitrary Riemann surfaces.
\end{remark}

In general, for $A=(a_1,...,a_r)\in \{1,3,...,2g-1\}^r$ and a function $G(u)$, 
$u=(u_1$, $u_3$, $\dots$, $u_{2g-1})^t$,
let
\bea
&&
\partial_A=\partial_{u_{a_1}}\cdots\partial_{u_{a_r}},
\hskip5mm
G_A(u)=\partial_A G(u),
\hskip5mm
\partial_{u_a}=\frac{\partial}{\partial u_a},
\non
\\
&&
|A|=a_1+\cdots+a_r.
\non
\ena
For $A=\emptyset$ we set $G_A(u)=G(u)$.

Some of the properties of $A^{(g)}_k$ are 

\begin{lemma}{\rm \cite{O2005}}\label{property-Agk}
(i) $\displaystyle |A^{(g)}_k|=\frac{1}{2}(g+1-k)(g-k)$.
\vskip2mm
\noindent
(ii) Let $\nu^{(g)}=(g,g-1,...,1)$. Then 
\bea
&&
\partial_{A^{(g)}_0} s_{\nu^{(g)}}(t)=s^{(g)}_0.
\non
\ena
\vskip2mm
\noindent
(iii) Suppose that $A\in \{1,3,...,2g-1\}^r$, $r\geq 1$ and $\nu$ be a partition.
If $|A|<|\nu|$ then $(\partial_A s_\nu)(0)=0$.
\end{lemma}
Notice that the property (iii) follows from the fact that the Schur function $s_{\nu}(t)$ is 
a weight homogeneous polynomial with the weight $|\nu|$. We include it in the lemma 
for the sake of later use.

The following theorem is proved in \cite{O2005}.

\begin{theorem} {\rm \cite{O2005}}\label{O-nonvanish}
$\theta[-K]_{A_0^{(g)}}(0|T^{(g)})\neq 0$.
\end{theorem}

\begin{remark}  Lemma \ref{property-Agk} and 
Theorem \ref{O-nonvanish} have been generalized to the case of $(n,s)$ curves in \cite {NY2012} and 
to the case of arbitrary Riemann surfaces in \cite{N2016}.
\end{remark}

\begin{definition}\label{def-sigma}
The fundamental sigma function of the hyperelliptic curve $C$ is defined for the data 
$(C,\infty, \{\rmd u_{2i-1}^{(g)}\}, \widehat{\Omega}^{(g)})$ by
\bea
&&
\sigma^{(g)}(u)=
\exp\left(\frac{1}{2}u^t \eta^{(g)}(\omega^{(g)})^{-1} u\right)
\frac{\theta[-K]((2\omega^{(g)})^{-1}u|T^{(g)})}{{\mathrm{s}}_0^{(g)}\theta[-K]_{A_0^{(g)}}(0|T^{(g)})},
\label{def-sigma-1}
\ena
where $u=(u_1,u_3,...,u_{2g-1})^t$.
\end{definition}

In the remaining part of the paper we call the fundamental sigma function simply the sigma function
for the sake of simplicity.

Similarly to Riemann's theta function the sigma function is a quasi-periodic function.

\begin{lemma} \cite{K1,K2,BEL1}
For $m_1,m_2\in{\mathbb Z}^g$ the sigma function satisfies the following relation,
\bea
&&
\sigma^{(g)}(u+\omega^{(g)} m_1+\omega'{}^{(g)} m_2)
\non
\\
&&
=(-1)^{m_1^tm_2+2(\kappa^tm_1-\kappa'{}^tm_2)}
\non
\\
&&
\times
\exp\left( 2(\eta^{(g)} m_1+\eta'{}^{(g)} m_2){}^t ( u+\omega^{(g)} m_1+\omega'{}^{(g)} m_2 ) \right)
\sigma^{(g)}(u).
\label{periodicity}
\ena
\end{lemma}

Assign the weight $i$ to $u_i$.
Then the sigma function has several nice properties compared with Riemann's theta function.

\begin{theorem}\label{Klein}\cite{K1,K2,BEL1999,N2010-a}
\vskip2mm
\noindent
(i) The function  $\sigma^{(g)}(u)$ does not depend on the choice of a canonical homology basis.
\vskip2mm
\noindent
(ii) Taylor coefficients of $\sigma^{(g)}(u)$ are weighted homogeneous polynomials of $\{\mu_i\}$ with the coefficients in 
${\mathbb Q}$.
\vskip2mm
\noindent
(iii) The following expansion is valid,
\bea
&&
\sigma^{(g)}(u)=s_{\nu^{(g)}}(u)+(\text{terms with weights $>|\nu|$}),
\non
\ena
where $\nu^{(g)}=(g,g-1,...,1)$.
\end{theorem}

\begin{remark} 
(i) Due to (i) of Theorem \ref{Klein} the choice of a canonical homology basis 
is not included in the data to define the sigma function in Definition \ref{def-sigma}.
The property (i) signifies that the sigma function is modular invariant, since 
the change of a canonical homology basis is described by the action of the 
symplectic group ${\rm Sp}(2g,{\mathbb Z})$. 
\vskip2mm
\noindent
(ii) The sigma function for a hyperelliptic curve had been introduced by Klein \cite{K1,K2}.
It coincides with the Weierstrass sigma function if $g=1$ and the equation of $C$ is given in the 
Weierstarss canonical form.
Theorem \ref{Klein} is essentially due to Klein. But the relation with Schur functions 
was discovered in \cite{BEL1999} and proved in \cite{N2010-a}. The definition of sigma functions 
are generalized to $(n,s)$ curves in \cite{BEL1999}, to telescopic curves in \cite{Ayano} and to 
arbitrary Riemann surfaces in \cite{KS}. The properties of sigma functions corresponding to 
Theorem \ref{Klein} had been established in \cite{N2010-a,N2010} for $(n,s)$ curves, in 
\cite{Ayano} for telescopic curves, and the properties corresponding to  (i), (iii) of 
Theorem \ref{Klein} were proved in \cite{N2016} for any Riemann surface.
\end{remark}

\subsection{Tau function corresponding to the frame $\tilde{X}$}
In this section we recall the representation of the tau function corresponding to 
the point $z^g\tilde{W}$ of UGM in terms of the fundamental sigma 
function \cite{N2010}.

Define $b_{ij}^{(g)}$, $\widehat{q}_{ij}^{(g)}$ and $c_i^{(g)}$ by the expansions at $\infty$,
\bea
\rmd u_{2i-1}^{(g)}&=&\left(\sum_{j=1}^\infty b_{ij}^{(g)}z^{j-1}\right)\rmd z,
\non
\\
\widehat{\Omega}^{(g)}(p_1,p_2)&=&\left(\frac{1}{(z_1-z_2)^2}+\sum_{i,j=1}^\infty \widehat{q}_{ij}^{(g)}
z_1^{i-1}z_2^{j-1}
\right)\rmd z_1 \rmd z_2,
\non
\\
\log \left( z^{-g+1}\sqrt{ \frac{\rmd u_{2g-1}^{(g)}}{\rmd z} }  \right)&=&\sum_{i=1}^\infty c_i^{(g)}
\frac{z^i}{i}.
\non
\ena
Using
\bea
&&
\rmd u_{2i-1}^{(g)}=\frac{z^{2i-2}\rmd z}{F(z)},
\non
\ena
$c_i^{(g)}$ can be expressed explicitly by $\{\alpha_j\}$ as follows.
We have 
\bea
\log \left( z^{-g+1}\sqrt{ \frac{\rmd u_{2g-1}^{(g)} }{\rmd z} }  \right)&=&
-\frac{1}{2}\log F(z) 
\non
\\
&=&
-\frac{1}{4}\sum_{j=1}^{2g+1} \log(1-\alpha_{j}z^2)
\non
\\
&=&
\sum_{i=1}^\infty \left(\frac{1}{2}\sum_{j=1}^{2g+1} \alpha_j^i\right)\frac{z^{2i}}{2i}.
\non
\ena
Thus
\bea
&&
c_i^{(g)}=\left\{
\begin{array}{ll}
0& \text{if $i$ is odd}\\ \displaystyle
\frac{1}{2}\sum_{j=1}^{2g+1} \alpha_j^{i/2} &\text{if $i$ is even.}\\
\end{array}
\right.
\label{ci}
\ena

Define 
\bea
&&
B^{(g)}=(b_{ij}^{(g)})_{1\leq i\leq g, j\geq 1},
\quad
\widehat{q}^{(g)}(t)=\sum_{i,j=1}^\infty \widehat{q}_{ij}^{(g)}t_it_j.
\label{Bg-qg}
\ena

Then

\begin{theorem}\label{tau-sigma-1} {\rm \cite{N2010}}
The tau function corresponding to the frame $\tilde{X}$ defined by (\ref{frame-tilde-X}) has the following expression
\bea
&&
\tau(t;\tilde{X})=\exp\left(-\sum_{i=1}^\infty c_{2i}^{(g)} t_{2i}+\frac{1}{2}\widehat{q}^{(g)}(t)\right)\sigma^{(g)}(B^{(g)}t).
\label{tau-nonsingular}
\ena
\end{theorem}

\begin{remark}
 In \cite{N2010} this theorem is proved for any $(n,s)$ curve.
\end{remark}

\subsection{Tau function corresponding to the frames $Y^\pm$}
In this section we derive the sigma function expression of the tau function 
corresponding to the frames $Y^\pm$ of UGM.
In \S3.3 $p_\pm=(\alpha,\pm y_0)$ was introduced as points on the curve $C'$ of genus $g-1$.
 In this section $p_\pm$ denote points of $C$  except in Theorem \ref{tau-Y},
where $C'$ is considered, for the sake of notational simplicity

For $p\in C$ set 
\bea
u^{(g)}_j(p)=\int_\infty^p \rmd u_j^{(g)},
\quad
u^{(g)}(p)=\big(u^{(g)}_1(p), u^{(g)}_3(p), \ldots, u^{(g)}_{2g-1}(p)\big)^t.
\label{abel-jacobi}
\ena
We omit the superscript $(g)$ of $u^{(g)}_j(p)$, $u^{(g)}(p)$ and denote them by $u_j(p)$, $u(p)$ respectively 
if there is no fear of confusion.


\begin{prop}\label{sigma-vanishing}{\rm \cite{O2005}}\label{sigma-expansion-1}
Let $1\leq k\leq g$.
\vskip2mm
\noindent
(i) Let $A\subset\{1,3,...,2g-1\}^r$, $r\geq 1$. If $|A|<|A^{(g)}_k|$ then 
\bea
&&
\sigma^{(g)}_A(u(p_1)+\cdots+u(p_k))=0,
\non
\ena
for all $p_1,...,p_k\in C$. 
\vskip2mm
\noindent
(ii) As a function of $p_1,...,p_k\in C$, $(\sigma^{(g)})_{A_k^{(g)}}(u(p_1)+\cdots+u(p_k))$ is not identically zero. 
\vskip2mm
\noindent
(iii) The following expansion in $z_k=z(p_k)$ holds,
\bea
&&
\sigma^{(g)}_{A_k^{(g)}}(u(p_1)+\cdots+u(p_k))
\non
\\
&&
\hskip20mm
={\mathrm{s}}_{k,k-1}^{(g)}\sigma^{(g)}_{A_{k-1}^{(g)}}(u(p_1)+\cdots+u(p_{k-1}))
z_k^{g+1-k}+O(z_k^{g +2-k}),
\label{sigma-expansion-2}
\ena
where ${\mathrm{s}}_{k,k-1}^{(g)}={\mathrm{s}}_k^{(g)}{\mathrm{s}}_{k-1}^{(g)}$ and, for $k=1$, 
$(\sigma^{(g)})_{A_{k-1}^{(g)}}(u(p_1)+\cdots+u(p_{k-1}))$ should be understood as $1$.
\end{prop}

\begin{remark} This proposition is generalized to the case of $(n,s)$ curves in \cite{NY2012}.
In \cite{N2016} the equivalent assertions to this proposition have been proved for 
the Riemann theta function of any Riemann surface.
\end{remark}

Since $p_+$ is the image of $p_-$ by the hyperelliptic involution $(x,y)\rightarrow (x,-y)$,

$$
u(p_{-})=-u(p_{+}).
$$

Therefore, we have, by the case $k=2$ of Proposition  \ref{sigma-expansion-1}  (iii), 
\bea
\sigma^{(g)}_{A_2^{(g)}}(u(p)-u(p_\pm))&=&\sigma^{(g)}_{A_2^{(g)}}(u(p)+u(p_\mp))
\non
\\
&=&
{\mathrm{s}}_{2,1}^{(g)}\sigma^{(g)}_{A_1^{(g)}}(u(p_{\mp}))z^{g-1}+O(z^{g})
\non
\\
&=&
{\mathrm{s}}_{2,1}^{(g)}\sigma^{(g)}_{A_1^{(g)}}(-u(p_{\pm}))z^{g-1}+O(z^{g}),
\label{expans-1}
\ena
where $z=z(p)$.

\begin{prop}{\rm \cite{O2005}}\label{nonvanish-1}
$\displaystyle{\sigma^{(g)}_{A_1^{(g)}}(-u(p_\pm))\neq 0.}$
\end{prop}
\vskip2mm
\noindent
{\it Proof.}  This proposition had been proved in \cite{O2005}. Since we use s similar 
argument to that in the proof of this proposition later, we give a proof here.

Set
$$
f(p)=\sigma^{(g)}_{A_1^{(g)}}(u(p)).
$$
By the case $k=1$ of Proposition \ref{sigma-vanishing} (ii), (iii) it does not vanish identically as a 
function of $p$ and has the expansion of the form 
\bea
&&
f(p)=s^{(g)}_{1,0} z^g+O(z^{g+1}).
\non
\ena
Therefore $f(p)$ has a zero of order $g$ at $\infty$. Then the proposition follows from (ii) of 
the following proposition, since $p_\pm\neq \infty$.

\begin{prop}\label{fp-property}
\vskip2mm
\noindent
(i) For a cycle $\gamma=\sum_{i=1}^g (m_{1,i}\mathfrak{a}_i+m_{2,i}\mathfrak{b}_i)$, 
$m_{1,i}, m_{2,i}\in {\mathbb Z}$,  let $f(\gamma p)$ 
denote the analytic continuation of $f(p)$ along $\gamma$. Then 
\bea
&&
f(\gamma p)=(-1)^{m_1^tm_2+2(\kappa^tm_1-\kappa'{}^tm_2)}
\non
\\
&&
\times
\exp\left( 2(\eta^{(g)} m_1+\eta'{}^{(g)} m_2){}^t (u(p)+\omega^{(g)} m_1+\omega'{}^{(g)} m_2 ) \right)
f(p),
\label{periodicity-2}
\ena
where $m_i=(m_{i,1},...,m_{i,g}){}^t$.
\vskip2mm
\noindent
(ii) The function $f(p)$ has precisely $g$ zeros on $C$.
\end{prop}

\begin{remark}
Since $f(p)$ is a composition of the Abel-Jacobi map and the analytic function 
$\sigma^{(g)}_{A^{(g)}_1}(u)$ on 
${\mathbb C}^g$, it is 
a multi-valued analytic function on $C$.
The property (i) of Proposition \ref{fp-property} means that $f(p)$ is in fact a quasi-periodic function. 
Then  whether $f(p)$ becomes zero or not 
does not depend on the choice of a branch of $f(p)$. Therefore the assertion of 
Proposition \ref{fp-property}, (ii) makes sense.
\end{remark}
\vskip3mm

\noindent
{\it Proof of Proposition \ref{fp-property}}
\par
\noindent
(i) Apply the derivative $\partial_{A^{(g)}_1}$ to both hand sides of (\ref{periodicity}) and 
set $\displaystyle u=u(p)$. Then the property (i) of Proposition \ref{sigma-vanishing}
implies the formula (\ref{periodicity-2}).
\vskip2mm
\noindent
(ii)  For the proof let us recall the following well-known property of Riemann's theta function (see 
 \cite{FK},\cite{Mum1} for example).

\begin{theorem}\label{zero-Riemann-theta}
Let $C$ be a compact Riemann surface of genus $g$, $p_0$ a point of $C$, 
$\{\mathfrak{a}_i,\mathfrak{b}_i|1\leq i\leq g\}$ a canonical homology basis,
$dv^{(g)}=(dv^{(g)}_1,...,dv^{(g)}_g){}^t$ the normalized basis of holomorphic one forms,
$\varepsilon,\varepsilon'\in {\mathbb R}^g$,
$v \in {\mathbb C}^g$ and $\displaystyle \tilde{f}(p)=\theta\begin{bmatrix} \varepsilon \\ 
\varepsilon' \end{bmatrix}(\int_{p_0}^p dv^{(g)}+v)$. If $\tilde{f}(p)$ does not vanish identically, then 
it has precisely $g$ zeros on $C$.
\end{theorem}
In the present case of the hyperelliptic curve $C$ we use the same notation $\{dv^{(g)}_j\}$ 
for the basis of the normailzed holomorphic one forms and set, for $p\in C$, 
\bea
&&
v^{(g)}_j(p)=\int_\infty^p dv^{(g)}_j,
\quad
v^{(g)}(p)=\big(v^{(g)}_1(p),v_2(p),\ldots,v^{(g)}_g(p)\big){}^t.
\non
\ena
Again the superscript $(g)$ is omitted if there is no fear of confusion.

The sigma function is defined by (\ref{def-sigma}) using the Riemann theta function.
Conversely Riemann's theta function is expressed by the sigma function as
\bea
&&
\theta[-K](v|T^{(g)})=C\exp\left(-\frac{1}{2}(2\omega^{(g)}v){}^t\eta^{(g)}(\omega^{(g)})^{-1}
(2\omega^{(g)}v)\right)\sigma^{(g)}(2\omega^{(g)}v),
\non
\ena
for some constant $C$.
With this in mind we define the function $F(p)$ by
\bea
&&
F(p)=\exp\left(-\frac{1}{2}(2\omega^{(g)}v(p)){}^t\eta^{(g)}(\omega^{(g)})^{-1}
(2\omega^{(g)}v(p))\right)\sigma^{(g)}_{A^{(g)}_1}\left(2\omega^{(g)}v(p)\right).
\non
\ena
Then $F(p)$ satisfies
\bea
F(\mathfrak{a}_j p)&=&\exp(2\pi i\kappa_j)F(p),
\non
\\
F(\mathfrak{b}_j p)&=&\exp(-2\pi i\kappa'_j-\pi i T^{(g)}_{j,j}-2\pi i v_j(p))F(p),
\non
\ena
where $\kappa_j$ is the $j$-th component of $\kappa$.
Then exactly the same proof can be applied to $F(p)$ as to $\tilde{f}(p)$ in 
Theorem \ref{zero-Riemann-theta}. Therefore $F(p)$ has precisely $g$ zeros on $C$.
Since
\bea
&&
2\omega^{(g)}v(p)=u(p),
\non
\ena
it follows that $\sigma^{(g)}_{A^{(g)}_1}(u(p))$ has $g$ zeros on $C$. $\Box$

By (\ref{expans-1}) and Proposition \ref{nonvanish-1}  the following definition of the numbers $d_j^{(g,\pm)}$, $j\geq 1$, makes sense,
\bea
&&
\log\left(z^{g-1}\frac{{\mathrm{s}}_{2,1}^{(g)}
\sigma^{(g)}_{A_1^{(g)}}(-u(p_\pm))}{\sigma^{(g)}_{A_2^{(g)}}(u(p)-u(p_\pm))}\right)-
\frac{1}{2}\widehat{q}^{(g)}([z])=\sum_{j=1}^\infty d^{(g,\pm)}_j\frac{z^j}{j},
\label{def-dj}
\ena
where, for $g=1$, we set $A^{(1)}_2=\emptyset$, ${\mathrm{s}}^{(1)}_{2,1}=1$.

Then
\begin{theorem}\label{tau-Y} The tau function corresponding to the frames $Y^\epsilon$, $\epsilon=\pm$,
defined by (\ref{frame-Y}) has the following expression
\bea
\tau(t;Y^\epsilon)&=&C_\epsilon
 \exp\left(\sum_{j=1}^\infty d_j^{(g-1,\epsilon)} t_j +\frac{1}{2} {\widehat q}^{(g-1)}(t)\right)
\sigma^{(g-1)}(B^{(g-1)} t-u(p_\epsilon)),
\label{tau-Y-1}
\\
C_\epsilon&=&\left({\mathrm{s}}_0^{(g-2)}\sigma^{(g-1)}_{A_0^{(g-2)}}(-u(p_\epsilon))\right)^{-1},
\non
\ena
where $d^{(g-1,\epsilon)}_j$ is defined by (\ref{def-dj}), $B^{(g-1)}$, $\widehat{q}^{(g-1)}$ 
are given by (\ref{Bg-qg}) and $A^{(g-2)}_0$, $s^{(g-2)}_0$, $p_\pm$ are given by 
(\ref{akg}), (\ref{signature}) , (\ref{p-pm}) respectively.
\end{theorem}
\vskip2mm
\noindent
{\it Proof.} Let us denote the right hand side of (\ref{tau-Y-1}) by $\tau_\epsilon(t)$. 
Notice first that $\tau_\epsilon(t)$ is a solution of the KP-hierarchy (\ref{KP-hierarchy}).
In fact it is proved in \cite{N2010,EEG} that $\tilde{\tau}(t)$ defined by
\bea
&&
\tilde{\tau}(t)=\exp\left(\frac{1}{2} {\widehat q}^{(g-1)}(t)\right)\sigma^{(g-1)}(B^{(g-1)} t+u)
\non
\ena
is a solution of the KP-hierarchy for any $u\in {\mathbb C}^g$. 
In general if $\tau_1(t)$ is a solution of (\ref{KP-hierarchy}) then 
$\tau_2(t):={\rm e}^{c_0+\sum_{i=1}^\infty c_i t_i}\tau_1(t)$ is a solution of 
 (\ref{KP-hierarchy}) for arbitrary constants $c_i$, $i\geq 0$.
The solution $\tau_2(t)$ is called the gauge transformation of $\tau_1(t)$. 
Since $\tau_\epsilon(t)$ is a gauge transformation of 
$\tilde{\tau}(t)$ with $\displaystyle u =-u^{(g-1)}(p_\epsilon)$, 
it is a  solution of the KP-hierarchy.

In order to determine the point of UGM corresponding to $\tau_\epsilon(t)$ according to (\ref{U-tau}) we compute the 
adjoint wave function (\ref{wave}) of $\tau_\epsilon(t)$,
\bea
&&
\Psi_\epsilon^\ast(t;z)=\frac{\tau_\epsilon(t+[z])}{\tau_\epsilon(t)}
\exp\left(-\sum_{i=1}^\infty t_i z^{-i}\right).
\label{wave-pm}
\ena

Let $\rmd \tilde{r}_i$ be the normalized differential of the second kind with a pole only at $\infty$ 
such that
\bea
&&
\rmd\tilde{r}_i=\rmd\left(\frac{1}{z^i}+O(1)\right) \quad \text{near $\infty$},
\non
\\
&&
\int_{\mathfrak{a}_j}\rmd\tilde{r}_i=0 \quad \text{for any $j$}.
\non
\ena
Set 
\bea
&&
\rmd\hat{r}_i=\rmd\tilde{r}_i+\sum_{k,l=1}^g b_{k,i}\left(\eta^{(g)}(\omega^{(g)})^{-1}\right)_{k,l} \rmd u_{2l-1}^{(g)}.
\label{def-dr-hat}
\ena
It is proved in Lemma 8 of \cite{N2010} that
\bea
&&
\rmd\hat{r}_i=\rmd\bigg(\frac{1}{z^i}-\sum_{j=1}^\infty \widehat{q}_{i,j}\frac{z^j}{j}\bigg).
\label{dr-hat}
\ena
By the definition of $B^{(g-1)}$ we have 
\bea
&&
B^{(g-1)}[z]=u^{(g-1)}(p),
\label{B-bracket-z}
\ena
for $z=z(p)$.
Using (\ref{def-dj}), (\ref{dr-hat}), (\ref{B-bracket-z}) we get
\bea
&&
\Psi_\epsilon^\ast(t;z)
\non
\\
&&
=z^{g-2}
\frac{{\mathrm{s}}_{2,1}^{(g-1)}
\sigma^{(g-1)}_{A_1^{(g-1)}}(-u(p_\epsilon))}{\sigma^{(g-1)}_{A_2^{(g-1)}}(
u(p)-u(p_\epsilon))}
\frac{\sigma^{(g-1)}(B^{(g-1)} t+u(p)-u(p_\epsilon))}{\sigma^{(g-1)}(B^{(g-1)} t-u(p_\epsilon))}
\non
\\
&&
\hskip5mm
\times
\exp\left(-\sum_{i=1}^\infty t_i \int^p \rmd\hat{r}_i \right),
\label{wave-formula}
\ena
where $\int^p \rmd\hat{r}_i$ is the indefinite integral without a constant term in the expansion in $z$.
We denote $z^{-(g-2)}\Psi_\epsilon^\ast(t;z)$ by $\Psi_\epsilon^\ast(p)$ which is a function of $p\in C'$.

\begin{lemma}\label{periodicity-psia}
(i) The functions $\Psi_\epsilon^\ast(p)$ are single valued on $C'$, that is, 
for any cycle $\gamma=\sum_{i=1}^g (m_{1,i}\mathfrak{a}_i+m_{2,i}\mathfrak{b}_i)$, 
$m_{1,i}, m_{2,i}\in {\mathbb Z}$, $\Psi_\epsilon^\ast(\gamma p)=\Psi_\epsilon^\ast(p)$.
\vskip2mm
\noindent
(ii) On $C'\backslash\{\infty\}$ $\Psi_\epsilon^\ast(p)$ is meromorphic with a simple pole only at $p_\epsilon$.
\end{lemma}
\noindent
{\it Proof.} (i) In a similar way to the proof of Proposition \ref{fp-property} (i) the following property can be 
 proved 
using (\ref{periodicity}) and Proposition \ref{sigma-vanishing} (i),
\bea
&&
\sigma^{(g-1)}_{A_2^{(g-1)}}(u(\gamma p)-u(p_\epsilon))=(-1)^{m_1^tm_2+2(\kappa^tm_1-\kappa'{}^tm_2)}
\non
\\
&&
\times
\exp\left( 2(\eta^{(g-1)} m_1+\eta'{}^{(g-1)} m_2){}^t ( u(p)-u(p_\epsilon)+
\omega^{(g-1)} m_1+\omega'{}^{(g-1)} m_2 ) \right)
\non
\\
&&
\times
\sigma^{(g-1)}_{A_2^{(g-1)}}(u(p)-u(p_\epsilon)),
\label{quasi-periodic-Ag2}
\ena
for $\gamma=\sum_{i=1}^{g-1} (m_{1,i}\mathfrak{a}_i+m_{2,i}\mathfrak{b}_i)$, where
 $m_i=(m_{i,1},...,m_{i,g-1}){}^t$.
 By the definition (\ref{def-dr-hat}) of  the differential $\rmd\hat{r}_i$ the following lemma can be proved.
 
 \begin{lemma}\label{period-dr-hat}\cite{N2010} 
 The periods of  $\rmd\hat{r}_i$ are given by
 \bea
 &&
 \int_{\mathfrak{a}_j}\rmd\hat{r}_i=\left((2\eta^{(g-1)}){}^t B^{(g-1)}\right)_{i,j},
 \hskip5mm
  \int_{\mathfrak{b}_j}\rmd\hat{r}_i=\left((2\eta'{}^{(g-1)}){}^t B^{(g-1)}\right)_{i,j},
 \ena
 \end{lemma}
 
 Using (\ref{periodicity}), (\ref{quasi-periodic-Ag2}), Lemma \ref{period-dr-hat} we easily have
 $\Psi^\ast_\epsilon(\gamma p)=\Psi^\ast_\epsilon(p)$.
 \vskip2mm
 \noindent

By Lemma \ref{periodicity-psia}  the expansion coefficients of $z^{-(g-2)}\tau_\pm(t)\Psi^\ast_\epsilon(t;z)$
in $t_1,t_2,...$ is contained in $W_\epsilon$ defined by (\ref{W-pm}).
By (\ref{U-tau}) the point $U_\epsilon$ of UGM corresponding to $\tau_\epsilon(t)$ is contained in $W_\epsilon$.

\begin{lemma}\label{comparison-points} If two points $U_1$, $U_2$ of UGM satisfy the inclusion relation $U_1\subset U_2$ then 
$U_1=U_2$.
\end{lemma}
\vskip2mm
\noindent
{\it Proof.} By $U_1\subset U_2$ we have
\bea
&&
{\rm Ker}(\pi|_{U_1}) \subset {\rm Ker}(\pi|_{U_2}),
\hskip10mm
{\rm Im}(\pi|_{U_1}) \subset {\rm Im}(\pi|_{U_2}),
\label{ineq-0}
\ena
and consequently 
\bea
\dim {\rm Ker}(\pi|_{U_1}) &\leq& \dim {\rm Ker}(\pi|_{U_2}),
\label{ineq-1}
\\
\dim {\rm Coker}(\pi|_{U_1}) &\geq& \dim {\rm Coker}(\pi|_{U_2}).
\label{ineq-2}
\ena
We show that these inequalities are actually equalities. 
In fact, if the strict inequality holds in (\ref{ineq-1}), for example,
then
\bea
&&
\dim {\rm Ker}(\pi|_{U_1}) <\dim {\rm Ker}(\pi|_{U_2})=\dim {\rm Coker}(\pi|_{U_2})\leq 
\dim {\rm Coker}(\pi|_{U_1})
\non
\ena
since $U_2\in {\rm UGM}$. But it contradicts $U_1\in {\rm UGM}$. 

Now suppose that $U_1\neq U_2$. Then there exists $v\in U_2$ such that $v\notin U_1$.
If $\pi(v)\neq 0$ then ${\rm Im}(\pi|_{U_1}) \subsetneq {\rm Im}(\pi|_{U_2})$ by (\ref{ineq-0}).
Therefore $\dim {\rm Coker}(\pi|_{U_1}) > \dim {\rm Coker}(\pi|_{U_2})$ which is impossible
as we have just shown. If $\pi(v)=0$, then ${\rm Ker}(\pi|_{U_1}) \neq {\rm Ker}(\pi|_{U_2})$ which 
is also impossible. Thus $U_1=U_2$. $\Box$

By Lemma \ref{comparison-points} we have $U_\epsilon=W_\epsilon$. Then there is a constant $c_\epsilon$ 
such that
\bea
&&
\tau_\epsilon(t)=c_\epsilon\tau(t;Y^\epsilon).
\label{const-tau}
\ena
We show that $c_\epsilon=1$ by comparing the expansion of tau functions.

By computing the Pl\"ucker coordinates of  $Y^\epsilon$ we easily see that the following expansion 
holds,
\bea
&&
\tau(t;Y^\epsilon)=s_{\nu^{(g-2)}}(t)+(\text{higher weight terms}).
\label{expansion-tau-Ypm}
\ena

The following proposition is proved in \cite{N2016}.

\begin{prop}{\rm \cite{N2016}}\label{expans-one-point-1}
Suppose that $q\in C\backslash\{\infty\}$. Then 
\bea
&&
\sigma^{(g-1)}(u+u(q))=c(q) s_{\nu^{(g-2)}}(t)|_{t_{2i-1}=u_{2i-1}, i\geq 1}+(\text{higher weight terms}),
\label{expans-one-point-2}
\ena
for some constant $c(q)\neq 0$.
\end{prop}
Notice that $s_{\nu^{(g-2)}}(t)$ depends only on $t_1,t_3,...,t_{2g-3}$.

\begin{remark} In \cite{N2016} the expansion of the form (\ref{expans-one-point-2}) is proved 
for the sigma function of an arbitrary Riemann surface.
\end{remark}

Apply $\partial_{A^{(g-2)}_0}$ to (\ref{expans-one-point-2}) and set $u=0$. Then 
By Lemma \ref{property-Agk} (ii), (iii) we have
\bea
&&
\sigma_{A^{(g-2)}_0}^{(g-1)}(u(q))=c(q) s^{(g-2)}_0.
\non
\ena
Therefore
\bea
&&
c(q)=s^{(g-2)}_0\partial_{A^{(g-2)}_0}\sigma^{(g-1)}(u(q)).
\non
\ena
It follows that
\bea
&&
c(-p_\epsilon)=C_\epsilon^{-1}.
\label{cpm}
\ena

By Proposition \ref{expans-one-point-1} and (\ref{cpm}) the expansion of $\tau_\epsilon(t)$
takes the form
\bea
&&
\tau_\epsilon(t)=s_{\nu^{(g-2)}}(t)+(\text{higher weight terms}).
\label{tau-epsilon}
\ena
Comparing (\ref{expansion-tau-Ypm}) and (\ref{tau-epsilon}) we get $c_\epsilon=1$ 
in (\ref{const-tau}). Thus Theorem \ref{tau-Y} is proved. $\Box$


\section{Degeneration of sigma function}
In this section we derive the formula for the degenerate sigma function by substituting the 
sigma function formula of the tau functions to the relation (\ref{decomposition-tau}).

\subsection{Relation of sigma functions}
Recall that $\tau(t;\tilde{X})$ is expressed by the sigma function as in Theorem \ref{tau-sigma-1}
and $\tau(t;\tilde{X}_0)$ is a degenerate limit of $\tau(t;\tilde{X})$. Since the point of UGM corresponding to 
$X$ is a gauge transformation of that of the frame $\tilde{X}_0$, $\tau(t;X)$ can be expressed by 
$\tau(t;\tilde{X}_0)$. We first derive it.

Expand 
\bea
&&
\log (1-\alpha z^2)^{-1}=\sum_{j=1}^\infty (2\alpha^j)\frac{z^{2j}}{2j}.
\non
\ena
By (\ref{tau-X}) 
\bea
&&
\tau(t;X)=\exp \bigg(2\sum_{j=1}^\infty \alpha^j t_{2j} \bigg)\tau(t;{\tilde X}_0).
\label{degenerate-tau}
\ena
We denote the limit of $\sigma^{(g)}(u)$ by $\sigma_g^{(g)0}(u)$ etc. by putting the upper index $0$ to each quantity. Then, by taking the limit of (\ref{tau-nonsingular}) we have
\bea
&&
\tau(t;X)=\exp\left(2\sum_{j=1}^\infty \alpha^j t_{2j}-\sum_{i=1}^\infty c_{2i}^{(g)0} t_{2i}+
\frac{1}{2}\widehat{q}^{(g)0}(t)\right)\sigma^{(g)0}(B^{(g)0}t).
\label{tau-singular}
\ena

Substitute (\ref{tau-Y-1}) and (\ref{tau-singular}) to (\ref{decomposition-tau}). We get

\begin{theorem}\label{degeneration-formula-0} 
Let $\sigma^{(g)}(u)$ be the sigma function defined by (\ref{def-sigma-1}) and $\sigma^{(g)0}(u)$ the limit of 
$\sigma^{(g)}(u)$ in the limit taking $\alpha_{2g+1}$, $\alpha_{2g}$ to $\alpha$ in the curve $C$ 
defined by (\ref{curve-1}). Then 
\bea
&&
\sigma^{(g)0}(B^{(g)0} t)=
\frac{(-1)^{g-1} }{2y_0}\mathrm{s}_0^{(g-2)} 
\exp \left(-2\sum_{j=1}^\infty \alpha^j t_{2j}+\sum_{i=1}^\infty c_{2i}^{(g)0} t_{2i}-
\frac{1}{2}\widehat{q}^{(g)0}(t)
+\frac{1}{2}\widehat{q}^{(g-1)}(t)\right)
\non
\\
&&
\times
\sum_{\epsilon=\pm}\epsilon
\exp\bigg(\sum_{j=1}^\infty d_j^{(g-1,\epsilon)} t_j\bigg)\frac{\sigma^{(g-1)}(B^{(g-1)} t-u(p_\epsilon))}
{\sigma^{(g-1)}_{A_0^{(g-2)}}(-u(p_\epsilon))},
\label{degeneration-formula-1}
\ena
where $\pm$ are identified with $\pm1$, $c^{(g)}_{2i}$, $B^{(g)}$, $\widehat{q}^{(g)}(t)$, $d^{(g,\epsilon)}_j$ 
are defined by (\ref{ci}), (\ref{Bg-qg}), (\ref{def-dj}) and $c^{(g)0}_{2i}$, $B^{(g)0}$, $\widehat{q}^{(g)0}(t)$ are 
the limits of the corresponding quantities.
\end{theorem}
In this formula the relation between the argument in $\sigma^{(g)0}$ and that in $\sigma^{(g-1)}$ 
is not clear. It can be given as follows.

Set
\bea
&&
B^{(g)0} t=(u_1,u_3,...,u_{2g-1})^t,
\label{def-u-variable}
\ena
which, in terms of coordinates, is written as
\bea
&&
u_{2i-1}=\sum_{j=1}^\infty b^{(g)0}_{i,2j-1}t_{2j-1}.
\hskip5mm
1\leq i\leq g,
\non
\ena
Then

\begin{lemma}
The following equation holds,
\bea
&&
B^{(g-1)} t=(u_1-\alpha u_3,u_3-\alpha u_5,...,u_{2g-3}-\alpha u_{2g-1}).
\label{Bmatrix-relation}
\ena
\end{lemma}
\vskip2mm
\noindent
{\it Proof.} The equation (\ref{Bmatrix-relation}) is equivalent to
\bea
&&
\sum_{j=1}^\infty b^{(g-1)}_{i,2j-1}t_{2j-1}=u_{2i-1}-\alpha u_{2i+1}.
\label{Bmatrix-relation2}
\ena
Let us prove it.
We have 
\bea
\rmd u^{(g)0}_{2i-1}&=&\frac{z^{2i-2} \rmd z}{(1-\alpha z^2)F_0(z)}=
\left(\sum_{j=i}^\infty b^{(g)0}_{i,2j-1} z^{2j-2}\right)\rmd z,
\label{g-du}
\\
\rmd u^{(g-1)}_{2i-1}&=&\frac{z^{2i-2} \rmd z}{F_0(z)}=
\left(\sum_{j=i}^\infty b^{(g-1)}_{i,2j-1} z^{2j-2}\right)\rmd z.
\label{g-1-du}
\ena
Therefore, for $1\leq i\leq g-1$, we have
\bea
&&
\rmd u^{(g-1)}_{2i-1}=(1-\alpha z^2)\rmd u^{(g)0}_{2i-1},
\label{1stdiff-exp2}
\ena
which is equivalent to
\bea
&&
b^{(g-1)}_{i,2j-1}=b^{(g)0}_{i,2j-1}-\alpha b^{(g)0}_{i,2j-3},
\non
\ena
where we set $b^{(g)0}_{ij}=0$ if $j\leq 0$.
Multiplying $t_{2j-1}$ and summing up we get
\bea
\sum_{j=i}^\infty b^{(g-1)}_{i,2j-1}t_{2j-1}
&=&
\sum_{j=i}^\infty b^{(g)0}_{i,2j-1} t_{2j-1}-
\alpha \sum_{j=i+1}^\infty  b^{(g)0}_{i,2j-3}t_{2j-1}
\non
\\
&=&
u_{2i-1}-\alpha \sum_{j=i+1}^\infty  b^{(g)0}_{i,2j-3}t_{2j-1}.
\label{Bmatrix-relation3}
\ena
Set
\bea
&&
\frac{1}{F(z)}=
\frac{1}{(1-\alpha z^2)F_0(z)}=\sum_{j=0}^\infty \beta_{2j+1} z^{2j}.
\label{def-beta}
\ena
By (\ref{g-du}) we have
\bea
&&
b^{(g)0}_{i,2j-1}=\beta_{2j-2i+1},
\non
\ena
and
\bea
&&
u_{2i-1}=\sum_{j=i}^\infty b^{(g)0}_{i,2j-1} t_{2j-1}=\sum_{j=i}^\infty \beta_{2j-2i+1} t_{2j-1}.
\non
\ena
Therefore 
\bea
&&
\sum_{j=i+1}^\infty b^{(g)0}_{i,2j-3} t_{2j-1}=
\sum_{j=i+1}^\infty \beta_{2j-2(i+1)+1} t_{2j-1}=u_{2i+1}.
\non
\ena
Substituting this to (\ref{Bmatrix-relation3}) we get (\ref{Bmatrix-relation2}).
$\Box$

By (\ref{def-u-variable}) and (\ref{Bmatrix-relation}) the arguments of  
$\sigma^{(g)0}$ and $\sigma^{(g-1)}$ in Theorem \ref{degeneration-formula-0} 
is directly connected without the variable { $t$}. Let us study other parts in the formula 
{(}\ref{degeneration-formula-1}{).}

\subsection{The expression of $ t$ by $u$}
In this section we express the variables $t_i$ in terms of $\{u_j\}$.

Define
\bea
\begin{split}
&
\Lambda_n=\frac{1}{2n}\sum_{j=1}^{2g+1} \alpha_j^n
= \frac{1}{2n}\bigg(2\alpha^2+\sum_{j=1}^{2g-1}\alpha_j^n\bigg)
\\
&
\Lambda=(\Lambda_1,\Lambda_2,\Lambda_3,\ldots)\\
\end{split}\label{Lambda}
\ena

\begin{lemma}\label{beta-expression}
$
\beta_{2j+1}=s_{(j)}(\Lambda),
\hskip5mm
j\geq 0.
$
\end{lemma}
\vskip2mm
\noindent
{\it Proof.} 
We have
\bea
\frac{1}{F(z)}= \exp(-\log F(z))&=&\exp\left(-\frac{1}{2}\sum_{j=1}^{2g+1}\log(1-\alpha_j z^2)\right)
\non
\\
&=&
\exp\left(\sum_{n=1}^\infty\left(\frac{1}{2n}\sum_{j=1}^{2g+1} \alpha_j^n\right)z^{2n}\right)
\non
\\
&=&
\exp\left(\sum_{n=1}^\infty \Lambda_n z^{2n}\right)
\non
\\
&=&
\sum_{n=0}^\infty s_{(n)}(\Lambda) z^{2n}.
\non
\ena
\vskip10mm
$\Box$

Set ${t}_{j}=0$, $j\geq 2g$ and ${t}_{2j-1}=0$, 
$j\geq g+1$ in (\ref{def-u-variable}). Then
\bea
&&
\left(
\begin{array}{c}
u_1\\u_3\\\vdots\\u_{2g-1}
\end{array}
\right)
=\tilde{M}
\left(
\begin{array}{c}{
t}_1\\{t}_3\\\vdots\\{t}_{2g-1}
\end{array}
\right),
\hskip5mm
\tilde{M}=\left(
\begin{array}{ccccc}
1&\beta_3&\beta_5&\ldots&\beta_{2g-1}\\
0&1&\beta_3&\ldots&\beta_{2g-3}\\
0&\ddots&\ddots&\quad&\vdots\\
\vdots&\quad&\ddots&\ddots&\vdots\\
0&\ldots&\ldots&0&1\\
\end{array}
\right).
\label{utEq}
\ena
Set
\bea
&&
M=\tilde{M}^{-1}.
\label{def-M}
\ena
The matrix $M$ can be computed as the cofactor matrix. 

\begin{lemma}\label{x-by-u}
Equation \eqref{utEq} is solved by
\bea
&&
\left(
\begin{array}{c}
{t}_1\\{t}_3\\\vdots\\{t}_{2g-1}
\end{array}
\right)
=M
\left(
\begin{array}{c}
u_1\\u_3\\\vdots\\u_{2g-1}
\end{array}
\right),
\hskip5mm
M=\left(
\begin{array}{ccccc}
1&m_{1,2}&m_{1,3}&\ldots&m_{1,g}\\
0&1&m_{2,3}&\ldots&m_{2,g}\\
0&\ddots&\ddots&\quad&\vdots\\
\vdots&\quad&\ddots&\ddots&\vdots\\
0&\ldots&\ldots&0&1\\
\end{array}
\right),
\non
\ena
where, for $i<j$,
\bea
&&
m_{ij}=
(-1)^{i+j}\det\left(
\begin{array}{ccccc}
s_{(1)}&s_{(2)}&\ldots&\ldots&s_{(j-i)}\\
1&s_{(1)}&s_{(2)}&\ldots&s_{(j-i-1)}\\
0&\ddots&\ddots&\quad&\vdots\\
\vdots&\ddots&\ddots&\quad&\vdots\\
0&\ldots&\ldots&1&s_{(1)}\\
\end{array}
\right),
\quad
s_{(j)}=s_{(j)}(\Lambda).
\label{matrix-M}
\ena
\end{lemma}

\subsection{Difference of bidifferentials}
In this section we express the coefficients of the quadratic form 
$-\frac{1}{2}\widehat{q}^{(g)0}(t)+\frac{1}{2}\widehat{q}^{(g-1)}(t)$ 
in terms of symmetric functions of $\{\alpha_j\}_{j=1}^{2g-1}$.

Let
\bea
Q_{i,j}&=&- \widehat{q}_{i,j}^{(g)0}+\widehat{q}_{i,j}^{(g-1)},
\hskip5mm
Q=(Q_{2i-1,2j-1})_{1\leq i,j\leq g},
\label{def-Q}
\ena
and
\bea
f_{g}(x)&=&(x-\alpha)^2\prod_{j=1}^{2g-1}(x-\alpha_j)=\sum_{i=0}^{2g+1}\mu_{4g+2-2i} x^i,
\non
\\
f_{g-1}(x)&=&\prod_{j=1}^{2g-1}(x-\alpha_j)=\sum_{i=0}^{2g-1}{\tilde \mu}_{4g-2-2i} x^i.
\non
\ena
We set $\tilde{\mu}_{4g-2-2i}=0$ if $i<0$ or $i>2g-1$.
The relation
\bea
&&
f_g(x)=(x-\alpha)^2f_{g-1}(x)
\non
\ena
implies
\bea
&&
\mu_{4g+2-2i} = \tilde{\mu}_{4g+2-2i}-2\alpha \tilde{\mu}_{4g-2i}+\alpha^2 \tilde{\mu}_{4g-2-2i}.
\label{mu-relation}
\ena
Define
\bea
H_g(x_1,x_2)&=&\sum_{i=0}^g(x_1x_2)^i \big(2\mu_{4g+2-4i}+\mu_{4g-4i}(x_1+x_2)\big),
\non
\\
H_{g-1}(x_1,x_2)&=&\sum_{i=0}^{g-1}(x_1x_2)^i\big(2\tilde{\mu}_{4g-2-4i}+\tilde{\mu}_{4g-4-4i}(x_1+x_2)\big).
\non
\ena

By computation we have

\begin{lemma}\label{H-relation}
\bea
&&
H_g(x_1,x_2)=(x_1-\alpha)(x_2-\alpha)H_{g-1}(x_1,x_2)+\alpha(x_1-x_2)^2\sum_{i=0}^{g-1}
\tilde{\mu}_{4g-4-4i} x_1^ix_2^i.
\non
\ena
\end{lemma}

Set $x_i=z_i^{-2}$ and 
\bea
y^{(g-1)}_i=z_i^{-(2g-1)}F_0(z)
\hskip10mm
y^{(g)}_i=(x_i-\alpha)y^{(g-1)}_i=z_i^{-(2g+1)}F(z).
\non
\ena

Then
\bea
\widehat{\Omega}^{(g)0}(x_1,x_2)
&=&
\left(
\frac{1}{2(x_1-x_2)^2}
+\frac{H_g(x_1,x_2)}{4(x_1-x_2)^2 y_1^{(g)} y_2^{(g)}}
\right)\rmd x_1 \rmd x_2,
\non
\\
\widehat{\Omega}^{(g-1)}(x_1,x_2)
&=&
\left(
\frac{1}{2(x_1-x_2)^2}
+\frac{H_{g-1}(x_1,x_2)}{4(x_1-x_2)^2 y_1^{(g-1)} y_2^{(g-1)}}
\right)\rmd x_1 \rmd x_2.
\non
\ena

By Lemma \ref{H-relation}
\bea
-\widehat{\Omega}^{(g)0}(x_1,x_2)+\widehat{\Omega}^{(g-1)}(x_1,x_2)
&=&
-\alpha\frac{\sum_{i=0}^{g-1}\tilde{\mu}_{4g-4-4i} x_1^ix_2^i}{4y_1^{(g)} y_2^{(g)}} \rmd x_1 \rmd x_2
\non
\\
&=&
-\alpha\frac{\sum_{i=0}^{g-1}\tilde{\mu}_{4i} z_1^{2i} z_2^{2i}}{F(z_1) F(z_2)} \rmd z_1 \rmd z_2.
\label{omega-difference}
\ena
We have
\bea
&&
\frac{1}{F(z)}=\sum_{n=0}^\infty s_{(n)}(\Lambda)z^{2n},
\label{F-expand}
\ena
by (\ref{def-beta}) and Lemma \ref{beta-expression}.

Compute the expansion of  the right hand side of (\ref{omega-difference})  using (\ref{F-expand}) 
and compare the expansion coefficients. Then we get

\begin{prop}\label{Q-formula}
For $i,j\geq 0$
\bea
&&
Q_{2i+1,2j+1}
=
-\alpha \sum_{k=0}^{g-1} \tilde{\mu}_{4k} s_{(i-k)}s_{(j-k)},
\label{Q-formula-1}
\ena
where $s_{(j)}=s_{(j)}(\Lambda)$.
\end{prop}

\noindent
\begin{example}\label{Q-genus2} The case of $g=2$.
\bea
&&
Q_{1,1}=-\alpha,
\hskip5mm
Q_{1,3}=Q_{3,1}=-\alpha s_{(1)},
\hskip5mm
Q_{3,3}=-\alpha(s_{(1)}^2+\tilde{\mu}_4),
\non
\ena
where 
\bea
s_{(1)}=\alpha-\frac{1}{2}\tilde{\mu}_2.
\non
\ena
In this case $m_{1,2}=s_{(1)}$ and 
\bea
&&
t_1=u_1-s_{(1)}u_3,
\hskip10mm
t_3=u_3.
\non
\ena
Then
\bea
&&
Q_{1,1}t_1^2+2Q_{1,3}t_1t_3+Q_{3,3}t_3^2
=-\alpha(u_1^2+\tilde{\mu}_4 u_3^2).
\non
\ena
\end{example}

\subsection{Final formula for the degenerate sigma function}
Substituting the results in the previous subsections to the formula in Theorem \ref{degeneration-formula-0} 
we now derive the final formula for the degenerate sigma function.

Set
\bea
u&=&(u_1,u_3,...,u_{2g-1})^t,
\label{def-u}
\\
\tilde{u}&=&(u_1-\alpha u_3,u_3-\alpha u_5,...,u_{2g-3}-\alpha u_{2g-1})^t,
\label{def-tilde-u}
\\
d^{ (g-1,\pm) }&=& \big(d^{ (g-1,\pm) }_1, d^{ (g-1,\pm)}_3, \ldots,d^{(g-1,\pm)}_{2g-1}\big)^t.
\label{def-d}
\ena
Since $\widehat{q}^{(g)0}(t)$, $\widehat{q}^{(g-1)}(t)$ do not contain $t_{2j}$, $j\geq 1$,
 we set $t_{2j}=0$, $j\geq 1$ and $t_{2j-1}=0$, $j\geq g+1$  in (\ref{degeneration-formula-1}).
Then

\begin{theorem}\label{main} Let $C$ be the hyperelliptic curve $C$ given by (\ref{curve-1}), 
$\sigma^{(g)}(u)$  the sigma function of $C$ defined by (\ref{def-sigma-1})
and $\sigma^{(g)0}(u)$  the 
limit of $\sigma^{(g)}(u)$ when $\alpha_{2g+1}$, $\alpha_{2g}$ $\to$ $\alpha$.
Then  $\sigma^{(g)0}(u)$ is given by the following formula 
\bea
&&
\sigma^{(g)0}(u)=
\frac{(-1)^{g-1}}{2y_0}\mathrm{s}_0^{(g-2)}
\exp \left(\frac{1}{2}  u^t M^t Q M  u \right) 
\non
\\
&&
\hskip20mm
\times
\sum_{\epsilon=\pm} 
\epsilon \exp\left( d^{(g-1,\epsilon)}{}^t M u \right)
\frac{\sigma^{(g-1)}(\tilde{u}-u(p_\epsilon))}{\sigma^{(g-1)}_{A_0^{(g-2)}}(-u(p_\epsilon))},
\label{main-1}
\ena
where $s^{(g-2)}_0$ is given by (\ref{signature}), $M$ is given by (\ref{matrix-M}), $Q$ is given by (\ref{def-Q}), (\ref{Q-formula-1}),
$d^{(g-1,\epsilon)}$ is given by (\ref{def-d}), (\ref{def-dj}),  $y_0$, $p_\epsilon$ are given by (\ref{p-pm}) and 
$u(p)$ is the Abel-Jacobi map defined by (\ref{abel-jacobi}).
\end{theorem}

\section{The case of genus two}

In \cite{BL} the degeneration of the genus two hyperelliptic sigma 
function is studied using the linear differential equations satisfied by the sigma function. 
In this section we derive their formula by our approach.

\subsection{Problem and the strategy}
In \cite{BL} some special normalization is imposed on the defining equation of 
the hyperelliptic curve. By this reason we can not apply the results in this paper 
directly to the case considered in \cite{BL}. Here we first explain what is the problem and how 
we treat it.

Let $\alpha_1,...,\alpha_5$ be mutually distinct complex numbers, $a$ a complex number and 
\bea
&&
C_1: w^2=\prod_{j=1}^5(v-\alpha_j)=\sum_{j=0}^5 \mu_{10-2j} v^j,
\quad
\mu_0=1,\ \mu_2=0,
\label{2-eq1}
\\
&&
C_2: y^2=\prod_{j=1}^5(x-\tilde{\alpha}_j)=\sum_{j=0}^5 \tilde{\mu}_{10-2j} x^j,
\quad
\tilde{\alpha}_j = \alpha_j+\frac{2}{3}\alpha,
\label{2-eq2}
\ena
genus two hyperelliptic curves. Two curves $C_1$ and $C_2$ are isomorphic 
by
\bea
&&
(x,y)\longrightarrow (v,w)=(x-\frac{2}{3}\alpha,y).
\label{2-eq3}
\ena

Assume that $\alpha \neq \alpha_j$, $j=1,2,3$ and consider the degeneration $\alpha_4, \alpha_5\rightarrow \alpha$ 
of the curve $C_1$ to
\bea
&&
w^2=(v-\alpha)^2\prod_{j=1}^3(v-\alpha_j)=\sum_{j=0}^5 \mu_{10-2j}^0 v^j.
\label{2-eq4}
\ena
Correspondingly the equation (\ref{2-eq2}) degenerates to
\bea
&&
y^2=\bigg(x-\frac{5}{3}\alpha\bigg)^2\bigg(x^3-\frac{g_2}{4} x-\frac{g_3}{4}\bigg),
\label{2-eq5}
\ena
where we set
\bea
&&
g_2=-4(\tilde{\alpha}_1\tilde{\alpha}_2+\tilde{\alpha}_1\tilde{\alpha}_3+\tilde{\alpha}_2\tilde{\alpha}_3),
\quad
g_3=4\tilde{\alpha}_1\tilde{\alpha}_2\tilde{\alpha}_3.
\label{2-eq6}
\ena
Notice that the coefficient of $x^2$ becomes zero in the second factor of the right hand side 
of (\ref{2-eq5}). This is the reason why we consider the shift by $(2/3)\alpha$ in (\ref{2-eq2}).
Notice that, due to $\mu_2=0$,  the coefficients of $\prod_{j=1}^3(v-\alpha_j)$  in (\ref{2-eq4}) 
becomes $2\alpha$ which is not zero if $\alpha\neq 0$. 

In \cite{BL} the limit of the sigma function of $C_1$ is expressed by the sigma function 
of the elliptic curve
\bea
&&
C_3: y^2=x^3-\frac{g_2}{4}x-\frac{g_3}{4}.
\label{2-eq7}
\ena

Although $C_1$ and $C_2$ are isomorphic, their sigma functions are not the same (see 
Remark \ref{remark-last}).
So we can not apply our result directly to the case studied in \cite{BL}. However the idea to
use tau function can be applied. The strategy is as follows. We first show that the tau 
functions of $C_1$ and $C_2$ are the same. By Theorem \ref{tau-degeneration-rel} 
the limit of the tau function of $C_2$ is expressed by the tau functions of  $C_3$. 
We can not apply Theorem \ref{tau-sigma-1} to the tau functions of $C_1$, 
since the sigma function formula of the tau function depends on the choice of a local coordinate.
As we shall see, for $C_1$, we need to use the local coordinate $z$ at $\infty$ given by
\bea
&&
v=\frac{1}{z^2}-\frac{2}{3}\alpha.
\label{2-eq8}
\ena
While Theorem \ref{tau-sigma-1} is derived using  the local coordinate $z$
such that
\bea
&&
v=\frac{1}{z^2}.
\label{2-eq9}
\ena
However it is possible to make the sigma function formula of the tau function 
using the local coordinate satisfying (\ref{2-eq8})
in a completely similar way to the case of (\ref{2-eq9}).

\subsection{Tau function of $C_1$ and $C_2$}
In this section we show that the tau function of $C_1$ and that of $C_2$ coincide.

Take the local coordinate $z$ at $\infty$ of $C_2$ by
\bea
&&
x=\frac{1}{z^2},
\quad
y=\frac{1}{z^5}(1+O(z)).
\non
\ena
By (\ref{2-eq3}) $z$ becomes a local coordinate at $\infty$ of $C_1$ such that
\bea
&&
{v}=\frac{1}{z^2}-\frac{2}{3}\alpha,
\quad
w=\frac{1}{z^5}(1+O(z)).
\label{2-eq10}
\ena

By (\ref{basis-1}), (\ref{local-coord}) a basis of $H^0(C_2, \cal{O}(\ast\infty))$ is given by 
\bea
&&
z^{-2i},
\quad
z^{-5-2i}F(z),
\quad
i\geq 0,
\label{2-eq11}
\ena
where 
\bea
&&
F(z)=\left(\prod_{j=1}^5(1-\tilde{\alpha}_j z^2)\right)^{\frac{1}{2}}.
\non
\ena
A basis of $H^0(C_1,\cal{O}(\ast\infty))$ is given by
\bea
&&
{v}^i,
\quad
{v}^i w,
\quad,
i\geq 0,
\non
\ena
which, in terms of $z$, is written as
\bea
&&
\left(z^{-2}-\frac{2}{3} \alpha\right)^i,
\quad 
z^{-5} \left(z^{-2}-\frac{2}{3} \alpha\right)^i  F(z),
\quad
i\geq 0.
\label{2-eq12}
\ena

We easily have

\begin{lemma}\label{space-C1-C2}
The vector space generated by (\ref{2-eq11}) and that by (\ref{2-eq12}) coincide.
\end{lemma}

By Lemma \ref{space-C1-C2} $C_1$ and $C_2$ with the local coordinate $z$ determine 
the same frame $\tilde{X}$ given by (\ref{frame-tilde-X}). 
Therefore their tau functions are the same.
Consequently Theorem \ref{tau-degeneration-rel} holds for tau functions of $C_1$ and $C_3$.

\subsection{$\tau(t;X)$ by sigma function of $C_1$}
In this section we express $\tau(t;X)$, which is a gauge transformation of the degenerate 
limit of $\tau(t;\tilde{X})$, in terms of the degenerate limit of the sigma function of $C_1$.

Redefine $b_{i,j}^{(g)}$, $\widehat{q}_{i,j}^{(g)}$, $c_i^{(g)}$ by the expansion coefficients 
in terms of the local coordinate $z$ of (\ref{2-eq10}). 
Then it can be proved that Theorem \ref{tau-sigma-1}
holds without any change, where the sigma function in the right hand side is that defined 
from $C_1$. In order to compare the formula in \cite{BL} we express the limits of 
$b_{i,j}^{(g)}$, $\widehat{q}_{i,j}^{(g)}$ in terms of the quantities associated with the curve $C_3$.

Let $\wp(u)$ be the Weierstrass elliptic function. The points of $C_3$ are parametrized as
\bea
&&
(x,y)=\left(\wp(u),\frac{1}{2}\wp'(u)\right),
\quad
u\in \Comp.
\non
\ena
By definition
\bea
&&
p_\pm=\left(\frac{5}{3}\alpha,\pm y_0\right)\in C_3.
\non
\ena
We choose $a \in \Comp$ such that
\bea
p_\pm=\left(\wp(a),\frac{1}{2}\wp'(\pm a)\right),
\label{2-eq13}
\ena
which means 
\bea
&&
\alpha=\frac{3}{5}\wp( a), 
\quad
y_0=\frac{1}{2}\wp'( a).
\label{2-eq14}
\ena

For the curve $C_3$
\bea
&&
\rmd u^{(1)}_1=-\frac{\rmd x}{2y}.
\non
\ena
Then 
\bea
&&
u(p_\pm)=\int_\infty^{p_\pm} \rmd u^{(1)}_1=-\int_\infty^{p_\pm}\frac{\rmd x}{2y}=-(\pm  a)=\mp a.
\label{2-eq15}
\ena

In the following everything is computed in the limit $\alpha_4,\alpha_5\rightarrow \alpha$.
We have
\bea
\rmd u^{(2)0}_1&=&-\frac{v \rmd v}{2w}=\big(1+ \alpha z +O(z^4)\big)dz,
\non
\\
\rmd u^{(2)0}_3&=&-\frac{\rmd v}{2w}=\big(z^2+O(z^4)\big)dz.
\non
\ena
Thus
\bea
&&
B^{(2)0}=
\left[
\begin{array}{cccccc}
1&0&\alpha&0&\ast&\ldots\\
0&0&1&0&\ast&\ldots
\end{array}
\right].
\label{2-eq16}
\ena
By definition
\bea
\widehat{\Omega}^{(2)0}(p_1,p_2)
&=&\frac{2w_1w_2+\sum_{i=0}^2 v_1^i v_2^i 
\Big(2\mu_{10-4i}^{(2)0}+\mu_{8-4i}^{(2)0}(v_1+v_2)\Big)}{4(v_1-v_2)^2w_1w_2}
\rmd v_1 \rmd v_2
\non
\\
&=&
\left(
\frac{1}{(z_1-z_2)^2}+\sum_{i,j\geq 1}\widehat{q}^{(2)0}_{2i-1,2j-1}z_1^{2(i-1)}z_2^{2(i-1)}
\right)
\rmd z_1 \rmd z_2.
\non
\ena
Then 
\bea
&&
\widehat{q}_{1,1}^{(2)0}=\frac{1}{5}\wp( a),
\quad
\widehat{q}_{1,3}^{(2)0}=\widehat{q}_{3,1}^{(2)0}=\frac{9}{25}\wp( a)^2+\frac{g_3}{8},
\non
\\
&&
\widehat{q}_{3,3}^{(2)0}=-\frac{3 g_2}{20} \wp( a)+\frac{3g_3}{8} + \frac{63}{125}\wp( a)^3.
\label{2-eq17}
\ena

Similarly to (\ref{ci}) $c_{2i-1}^{(2)0}=0$, $i\geq 1$.
If 
\bea
&&
t_{2j}=t_{2j+3}=0, \quad j\geq 1,
\label{2-restriction}
\ena
we have
\bea
&&
B^{(2)0}t=
\left[
\begin{array}{c}
t_1+\alpha t_3\\
t_3
\end{array}
\right],
\label{2-eq18}
\\
&&
\widehat{q}^{(2)0}(t)=\sum_{i,j=1}^2 \widehat{q}_{2i-1,2j-1}^{(2)0}t_{2i-1}t_{2j-1}.
\label{2-eq19}
\ena

Therefore, taking (\ref{degenerate-tau}) into account, we have
\bea
&&
\tau\big((t_1,0,t_3,0,\ldots)^t;X\big)
=\exp\left(\frac{1}{2}\widehat{q}^{(2)0}(t)\right)\sigma^{(2)0}(t_1+ \alpha t_3,t_3).
\label{2-tau-X}
\ena

\subsection{Tau function $\tau(t;Y^\pm)$ by sigma function of $C_3$}
In this section we derive the expression of $\tau(t;Y^\pm)$ in terms of the sigma 
function of $C_3$.

Put $g=2$ in Theorem \ref{tau-Y} and use $s^{(0)}_0=1$, $A^{(0)}_0=\emptyset$. Then 
\bea
&&
\tau\big((t_1,0,t_3,0,\ldots)^t;Y^\pm\big)
\non
\\
&&
\hskip20mm
=
\exp \Bigg( \sum_{j=1,3} d_j^{(1,\pm)} t_j+\frac{1}{2}\widehat{q}^{(1)}(t)\Bigg)
\frac{\sigma(B^{(1)} t-u(p_\pm))}{\sigma(-u(p_\pm))},
\label{2-eq20}
\ena
where $\sigma(u)=\sigma^{(1)}(u)$ is the Weierstrass sigma function.

We have 
\bea
&&
\rmd u^{(1)}_1=-\frac{\rmd x}{2y}= \big(1+O(z^4)\big) \rmd z,
\non
\ena
and
\bea
&&
B^{(1)}=[1,0,0,0,\ast,\ldots].
\non
\\
&&
B^{(1)}t=t_1.
\label{2-eq21}
\ena

For the bidifferential of $C_3$ we have
\bea
&&
\widehat{\Omega}^{(1)}(p_1,p_2)
\non
\\
&=&
\frac{2y_1y_2+x_1x_2(x_1+x_2)-\frac{1}{4}g_2(x_1+x_2)-\frac{1}{2}g_3}{4(x_1-x_2)^2y_1y_2} \rmd x_1 \rmd x_2
\non
\\
&=&
\left(
\frac{1}{(z_1-z_2)^2}+\frac{g_2}{8}(z_1^2+z_2^2)+\frac{g_3}{8}(z_1^4+z_2^4+3z_1^2z_2^2)+\cdots
\right)\rmd z_1 \rmd z_2.
\non
\ena
Therefore
\bea
&&
\widehat{q}^{(1)}_{1,1}=0,
\quad
\widehat{q}^{(1)}_{1,3}=\widehat{q}^{(1)}_{3,1}=\frac{g_2}{8},
\quad
\widehat{q}^{(1)}_{3,3}=\frac{3 g_3}{8}.
\label{2-eq22}
\ena
By (\ref{def-dj}) and (\ref{2-eq15})
\bea
&&
\log\left(\frac{\sigma(\pm  a)}{\sigma(p\pm  a)}\right)
-\frac{1}{2}\widehat{q}^{(1)}([z]|[z])
=\sum_{j=1}^\infty d_j^{(1,\pm)}\frac{z^j}{j}.
\label{2-eq23}
\ena
Since $\widehat{q}^{(1)}([z]|[z])$ contains only even powers of $z$, $d^{(1,\pm)}_j$, $j=1,3$ are
the expansion coefficients of the first term of the left hand side of (\ref{2-eq23}).
By computation we have
\bea
&&
d_1^{(1,\pm)}=\mp \zeta( a), 
\quad
d_3^{(1,\pm)}=\pm \frac{1}{2} \wp'( a),
\label{2-eq24}
\ena
where $\zeta(u)$ is the Weierstrass zeta function.
Substituting (\ref{2-eq21}), (\ref{2-eq24}) and  (\ref{2-eq15}) into (\ref{2-eq20}) we have
\bea
&&
\tau\big((t_1,0,t_3,0,\ldots)^t;Y^\pm\big)
\non
\\
&&
\hskip20mm
=
\exp\left(\mp \zeta( a)t_1\pm \frac{1}{2}\wp'( a) t_3+\frac{1}{2}\widehat{q}^{(1)}(t)\right)
\frac{\sigma(t_1\pm  a)}{\sigma(\pm  a)}.
\label{2-tau-Y}
\ena

\subsection{Genus two degenerate formula}
In this section we derive the formula of \cite{BL} based on the results of previous 
subsections.

Substitute (\ref{2-tau-X}) and (\ref{2-tau-Y}) into (\ref{decomposition-tau}) in 
Theorem \ref{tau-degeneration-rel}, use (\ref{2-eq14}) and 
solve in $\sigma^{(2)0}$. Then we have
\bea
&&
\sigma^{(2)0}(t_1+ \alpha t_3,t_3)
\non
\\
&=&
\frac{-1}{\wp'( a)}
\exp\left(
\frac{1}{2}\left(-q^{(2)0}(t)+q^{(1)}(t)\right)
\right)
\non
\\
&&
\times
\sum_{\epsilon=\pm1}
\epsilon \exp\left(-\epsilon \zeta( a)t_1+\epsilon \frac{1}{2}\wp'( a) t_3\right)
\frac{\sigma(t_1+\epsilon  a)}{\sigma(\epsilon  a)}.
\label{2-eq25}
\ena

Set
\bea
&&
u_1=t_1+\alpha t_3.
\quad
u_3=t_3.
\label{2-eq26}
\ena
Solving in $x$ we have
\bea
&&
t_1=u_1-\alpha u_3,
\quad
t_3=u_3.
\label{2-eq27}
\ena
Then
\bea
\frac{1}{2} \big(-q^{(2)0}(t)+q^{(1)}(t)\big)
&=&
-\frac{1}{10}\wp( a)u_1^2-\frac{6}{25}\wp( a)^2u_1u_3
\non
\\
&&
\hskip5mm
+
\left(-\frac{9}{125}\wp( a)^3+\frac{3}{40}g_2\wp( a)\right)u_3^2,
\label{2-eq28}
\\
-\zeta( a)t_1+\frac{1}{2}\wp( a)t_3&=&
-\zeta( a)u_1+\left(\frac{3}{5}\zeta( a)\wp( a)+\frac{1}{2}\wp'( a)\right)u_3.
\label{2-eq29}
\ena

Substitute (\ref{2-eq27}), (\ref{2-eq28}), (\ref{2-eq29}) into (\ref{2-eq25}) we get, with
$u=(u_1,u_3)^t$, 
\bea
&&
\sigma^{(2)0}(u)=
\frac{-1}{\sigma( a)\wp'( a)}
\exp\left(
-\frac{3}{5}\wp( a)\left(\frac{1}{6}u_1^2+\frac{2}{5}\wp( a)u_1u_3+
\Big(\frac{3}{25}\wp( a)^2-\frac{g_2}{8}\Big)u_3^2 \right)\right)
\non
\\
&&
\hskip18mm
\times
\sum_{\epsilon=\pm1}
\exp\left(\epsilon \left(-\zeta( a)u_1
+ \Big(\frac{3}{5}\zeta( a)\wp( a)+ \frac{1}{2}\wp'( a) \Big)u_3
\right)
\right)
\non
\\
&&
\hskip30mm
\times
\sigma\left(u_1-\frac{3}{5}\wp( a) u_3+\epsilon  a\right),
\label{2-eq30}
\ena
which coincides with the formula in Theorem 1 of \cite{BL} up to sign.
The discrepancy of the sign is due to the different normalization of the sigma function.

\begin{remark}\label{remark-last} Recall that the left hand side of  (\ref{2-eq30}) is the degenerate sigma function of the curve $C_1$.
The formula of Theorem \ref{main} gives a formula for the degenerate sigma function of the curve $C_2$.
By Example \ref{Q-genus2} it is obtained in (\ref{2-eq30}) by replacing the exponent of the exponential in front of the sum symbol 
by 
\bea
&&
-\frac{5}{6}\alpha (u_1^2+\tilde{\mu}_4 u_3^2).
\non
\ena
It shows that the sigma functions of $C_1$ and $C_2$ do not coincide.
\end{remark}

\vskip10mm
\noindent
{\bf \large Acknowledegements}
\vskip2mm
\noindent
We thank the anonymous referees for careful reading the manuscript and for their comments
that helped us to significantly improve the manuscript.
This work was supported by JSPS  KAKENHI Grant Number JP15K04907.

\end{document}